\documentclass[conference]{IEEEtran}
\usepackage{cite}
\usepackage{amsmath,amssymb,amsfonts}
\usepackage{algorithmic}
\usepackage{graphicx}
\usepackage{textcomp}
\usepackage{xcolor}
\usepackage[hyphens]{url}
\usepackage{fancyhdr}
\usepackage{hyperref}
\usepackage{booktabs}
\usepackage{tabularx}
\usepackage{adjustbox}
\usepackage{makecell} % For line breaks within a cell
\usepackage{tikz}
\usepackage{multirow}
\PassOptionsToPackage{table,xcdraw}{xcolor} % Ensure xcolor is loaded with necessary options before any package can load it
\usepackage{array} % For extending column definitions
\usepackage{colortbl} % For coloring cells and rows
\usepackage{comment}
\usepackage{enumitem}
\usepackage{tcolorbox}
\usepackage{filecontents}
\def\BibTeX{{\rm B\kern-.05em{\sc i\kern-.025em b}\kern-.08em
    T\kern-.1667em\lower.7ex\hbox{E}\kern-.125emX}}
\begin{document}

\title{Exploring and Evaluating Real-world CXL: Use Cases and System Adoption
% \thanks{This work was partially supported by U.S. National Science Foundation (2104116 and 2316202) and the Chameleon Cloud.}
}

\author{
\IEEEauthorblockN{Xi Wang\textsuperscript{*}}
\IEEEauthorblockA{\textit{University of California, Merced}\\
Merced, CA, USA \\
swang166@ucmerced.edu}
\and
\IEEEauthorblockN{Jie Liu\textsuperscript{*}}
\IEEEauthorblockA{\textit{University of California, Merced}\\
Merced, CA, USA \\
jliu279@ucmerced.edu}
\and
\IEEEauthorblockN{Jianbo Wu}
\IEEEauthorblockA{\textit{University of California, Merced}\\
Merced, CA, USA \\
jwu323@ucmerced.edu}
\and
\IEEEauthorblockN{Shuangyan Yang}
\IEEEauthorblockA{\textit{University of California, Merced}\\
Merced, CA, USA \\
syang127@ucmerced.edu}
\and
\IEEEauthorblockN{Jie Ren}
\IEEEauthorblockA{\textit{William \& Mary}\\
Williamsburg, VA, USA \\
jren03@wm.edu}
\and
\IEEEauthorblockN{Bhanu Shankar}
\IEEEauthorblockA{\textit{MemVerge, Inc}\\
Milpitas, CA, USA \\
bhanu.shankar@memverge.com}
\and
\IEEEauthorblockN{Dong Li}
\IEEEauthorblockA{\textit{University of California, Merced}\\
Merced, CA, USA \\
dli35@ucmerced.edu}
}

\maketitle

\renewcommand{\thefootnote}{\fnsymbol{footnote}} % Use symbols for footnotes
\footnotetext{\textsuperscript{*}Co-first authors.}
\renewcommand{\thefootnote}{\arabic{footnote}} % Revert back to numbers if needed later

\pagestyle{plain}

\definecolor{ren}{rgb}{0, 0, 0}
\definecolor{dong}{RGB}{0,0,255}
\definecolor{check}{RGB}{0,0,0}
\definecolor{jie}{RGB}{0,0,0}
\definecolor{sherry}{RGB}{255,128,0}
\definecolor{lightgray}{gray}{0.9}
\definecolor{revision}{RGB}{0, 0, 0}

\newcommand*\circled[1]{\tikz[baseline=(char.base)]{
            \node[shape=circle,draw,inner sep=1pt] (char) {#1};}}

%%%%%%%%%%%%%%%%%%%%%%%%%%%%%%%%%%%%%%%%
%%%%%%%% -- PAPER CONTENT STARTS -- %%%%%%%%%

\begin{abstract}
  Compute eXpress Link (CXL) is emerging as a promising memory interface technology. %Compared with the traditional memory interface (such as DDR), CXL provides multiple benefits, such as scalable increasing of memory capacity and bandwidth and avoiding memory stranding. However, 
  %Because of the common unavailability of CXL devices, the performance of the CXL memory is largely unknown. 
  However, its performance characteristics remain largely unclear due to the limited availability of production hardware. 
  Key questions include: What are the use cases for the CXL memory? What are the impacts of the CXL memory on application performance? How to use the CXL memory in combination with existing memory components? \textcolor{check}{In this work, we study the performance of three genuine CXL memory-expansion cards from different vendors. We characterize the basic performance of the CXL memory, study how HPC applications and large language models (LLM) can benefit from the CXL memory, and study the interplay between memory tiering and page interleaving. We also propose a novel data object-level interleaving policy to match the interleaving policy with memory access patterns. Our findings reveal the challenges and opportunities of using the CXL memory.} 
  %\textcolor{red}{Submissions may have at most 12 pages of technical content, including all text, figures, tables, appendices, etc}
\end{abstract}

%\begin{IEEEkeywords}
%\textcolor{red}{Compute Express Link, High Performance Computing, Large Language Model, LLMs inference.}
%\end{IEEEkeywords}

\section{Introduction}
\label{sec:intro}
\begin{figure}[!t]
	\centering
	\includegraphics[width=0.9\linewidth]{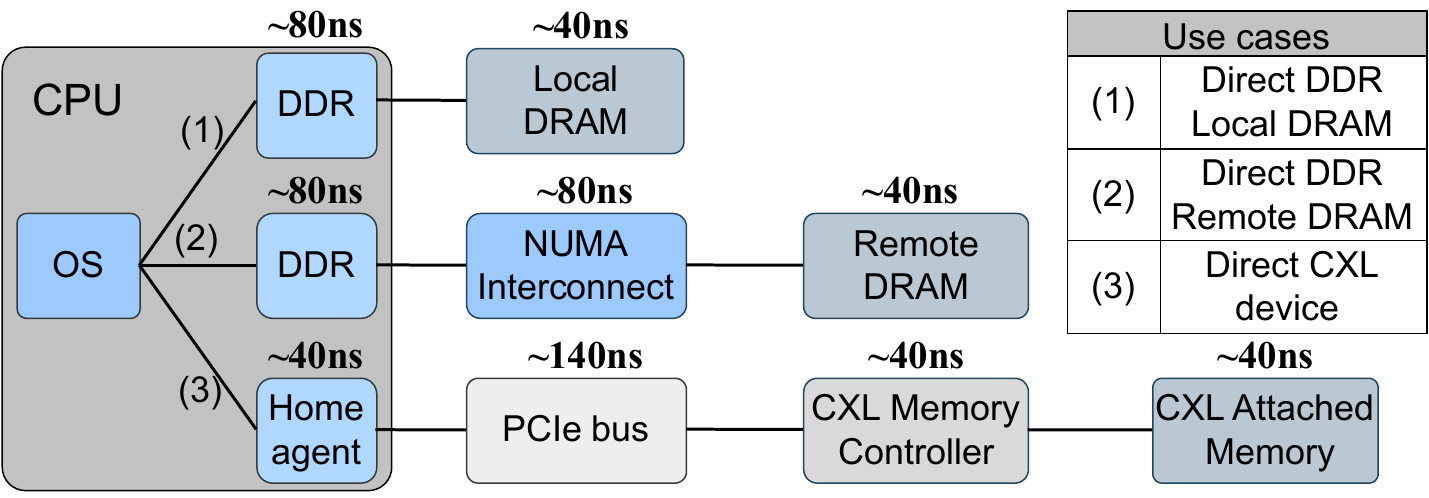}
        \vspace{-5pt}
	\caption{Breakdown of CXL memory access latency.}
	\centering
	\label{fig:latency_breakdown}
        \vspace{-20pt}
\end{figure}

Compute eXpress Link (CXL) is a promising memory interface technology. Based on the standard PCIe serial interface, CXL attaches memory to the CPU and appears as a CPU-less NUMA node. The CXL memory can be accessed in a cache-coherent fashion using load/store instructions. 
%%%Compared with the traditional memory interface technology (such as DDR), CXL provides multiple benefits, such as easily increasing memory capacity and bandwidth, and communicating with the attached memory in a cache-coherent fashion using traditional load/store instructions. 
However, CXL memory introduces longer memory access latency.  This longer latency comes from PCIe, %(e.g., \~ 40 ns in PCIe 4.0), 
CXL memory controller, and CXL home agent (HA) on the CPU.  Figure~\ref{fig:latency_breakdown} compares local NUMA, traditional remote NUMA, and CXL-based memory expansion in terms of memory latency. %%%It has been reported that the CXL adds about 70-150 $ns$ of extra latency over normal DRAM access~\cite{10.1145/3582016.3582063}.

Given the CXL performance, we face a series of questions: what are the use cases for the CXL memory? What are the impacts of real CXL memory on application performance? At the application level, how to use the CXL memory in combination with fast memory components (e.g., using uniform page-level interleaving vs. data object-level interleaving vs. memory binding)? %How do the existing page management systems perform in the real CXL hardware?  
This paper aims to discuss those questions, %characterize the performance of the CXL memory, 
and explore various paths to use the CXL memory. We study three genuine CXL memory expansion cards \textcolor{revision}{instead of using memory  simulation or emulation}. %%%%from different vendors. %%%%and have the following insights. 
%three genuine CXL-ready systems (\textcolor{dong}{including three CXL memory expansion cards from different memory vendors}), and have the following insights. %%%using ASIC-based hard CXL IPs.

\textbf{The CXL memory is a unique ``NUMA node''.}  The CXL memory appears as a CPU-less NUMA node. In the three systems we evaluate, the CXL memory appears as a two-hop-away NUMA node in terms of access latency. Depending on the memory vendors, the peak bandwidth of the CXL memory varies a lot, ranging from 9.8\% to 80.3\% of the peak bandwidth of local DRAM. %In addition, as we scale the number of threads to access the CXL memory, its memory bandwidth is easily saturated because of \textcolor{check}{single DRAM channel} (the saturation point is reached when the number of threads is only 4), while local DRAM (or LDRAM for short) and remote DRAM-based NUMA node (or RDRAM for short) have much better scalability. 
Moreover, as we increase the number of threads accessing the CXL memory, its bandwidth is quickly saturated due to \textcolor{ren}{the limited data transfer rate in CXL attached memory instead of PCIe bus — with saturation occurring when the number of threads reaches just four. }
In contrast, local DRAM (LDRAM) and remote DRAM-based NUMA nodes (RDRAM) exhibit much better scalability. 
This scaling difference between the CXL memory and DRAM highlights the importance of appropriately distributing memory accesses between them for high performance. Also, when the system is under heavy load, %we notice that LDRAM and RDRAM can show the similar latency as the CXL memory because of the contention on the memory controller (MC) or data path, which shows the potential of using CXL as LDRAM or RDRAM for latency-sensitive applications when the system is under heavy load. 
we observe that the latencies of LDRAM and RDRAM are similar to that of the CXL memory because of the contention on the memory controller (MC) or data path, which shows the potential of using CXL as LDRAM or RDRAM for latency-sensitive applications. %%%under heavy load.

\textbf{Using the CXL memory for large language models (LLM) faces challenges.} \textcolor{revision}{LLM can be  memory-consuming and have execution phases sensitive to memory bandwidth~\cite{kim2023squeezellm, atc21:zerooffload, patel2024splitwise}, hence having potentials to benefit from CXL.} We study the cases that use the CPU memory as an extension to the GPU memory to enable LLM training (using ZeRO-Offload~\cite{atc21:zerooffload}) and inference (using FlexGen~\cite{flexgen}) such that we can use less GPUs for LLM without the constraint of GPU memory capacity. This method has to frequently copy tensors between the GPU and CPU, and offloads certain computations from the CPU to the GPU to maximize GPU memory saving or reduce I/O offload. This method (named \textit{tensor offloading} in the rest of the paper) has been commonly studied and deployed in industry. 

We find that using tensor offloading based upon the CXL memory brings limited performance improvement. This is because the tensor copy between the CXL memory and GPU memory goes through a longer data path than memory accesses directly from CPU. Using CXL 1.1 on our platform, this data path is bottlenecked by the PCIe interconnect between CPU and GPU. As a result, adding CXL cannot show the bandwidth benefits for tensor offloading. In contrast, the computation offloaded to the CPU (i.e., the optimizer for LLM training and the attention computation for LLM inference), which tends to be bandwidth sensitive, can benefit from the extra bandwidth of CXL.
%suffers from the performance loss when using CXL. As a result, using the CXL memory for tensor offloading cannot show obvious performance improvement. 
In addition, the CXL memory increases memory capacity and allows us to use larger batch sizes for LLM inference, leading to throughput gains. \textcolor{check}{Our study is unprecedented, because of its practicality of using real CXL for GPU-based LLM, different from CPU-based study~\cite{eurosys24_cxl}.}
%%%Compared to using NVMe, using the CXL memory shows performance benefits, because of the large capacity of the CXL memory, we are able to use larger batch size for LLM inference, which leads to throughput gains.  %%%\textcolor{red}{(pending to add a couple of more sentences.)}

%\textbf{HPC applications can benefit a lot from the CXL memory for high performance while saving fast memory.} 
\textbf{\textcolor{check}{CXL can be used to save fast memory without causing performance loss, and using application semantics to guide page interleaving for CXL can maximize CXL benefits.}} 
%Previous work~\cite{micro23_cxl} reveals that complex applications (e.g., social network microservices) exhibiting $ms$-scale latency experience a marginal increase in tail latency even when most of pages are allocated to CXL memory. 
Our work goes beyond the existing work~\cite{micro23_cxl, eurosys24_cxl} that focuses on applications exhibiting $ms$-scale latency (e.g., social network microservices) or commercial workloads to study the potential of the CXL memory. We study a spectrum of HPC workloads, covering the most common and representative ``HPC dwarfs''~\cite{Asanovic+:TR06}. We reveal that some HPC applications \textcolor{check}{(such as CG and BT~\cite{NPB})} can tolerate the low bandwidth and high latency of the CXL memory under certain scales (the performance loss is less than 3.2\%, compared with LDRAM), because of their compute-intensive nature. %%%%%\textcolor{check}{Even for a latency-sensitive application, using CXL memory can bring high performance, transforming high bandwidth to the latency reduction.}
%%%%%We also observe that the bandwidth-sensitive and latency-sensitive applications respond differently to the CXL performance. %%\textcolor{red}{As we scale the number of threads, the bandwidth-sensitive application narrows down the performance difference between using CXL and not using CXL, because of the bandwidth benefit offered by CXL}, while the latency-sensitive application increases the difference. 

In addition, the page interleaving policy, embraced by the industry (e.g., Micron and AMD~\cite{micron_hetero_interleaving}, Astera Labs~\cite{asteralabs_hetero_interleaving}, and Samsung~\cite{samsung_hetero_interleaving}) as an application-transparent technique to integrate CXL with the existing memory components, provides opportunities to save LDRAM for HPC applications. When interleaving CXL and RDRAM,  we see minor performance difference from interleaving CXL and LDRAM for some applications. This is because the CXL memory dominates the memory performance, and the performance of other memory components has minor impact on the overall performance. 

To maximize the interleaving performance, we introduce a novel data object-level interleaving %\textcolor{sherry}{(or object-level interleaving for short)} 
policy.  
%\textcolor{sherry}{Different from the Linux uniform page-level interleaving (or uniform interleaving for short)  policy~\cite{micro23_cxl}}, 
Different from Linux uniform page-level interleaving~\cite{micro23_cxl}, this policy decides whether memory pages allocated to a data object should be interleaved between CXL and DRAM or allocated to LDRAM first (``LDRAM preferred''). This policy maximizes memory bandwidth (or minimizes latency) for data objects whose accesses favor high bandwidth (or low latency). %and minimizes the latency for data objects whose accesses favor low latency. 
This policy reduces LDRAM usage by \textcolor{check}{$32\%$} and outperforms the uniform interleaving policy (Linux default) by $65\%$ on average.

\textbf{Memory tiering solutions need to be improved.} Treating the CXL memory as a memory tier, existing work~\cite{autonuma, tiering0.8, eurosys24:mtm, 10.1145/3582016.3582063, Raybuck2021HeMemST, 10.1145/3600006.3613167} migrates pages between the CXL memory and fast memories based on page access frequency or recency (i.e., hotness). Those solutions are seldom studied with the real CXL memory, and \textit{how they interplay with the existing system (e.g., page interleaving) is largely unknown.} 

We find that the dynamic page migration in memory tiering are not integrated well with the static page interleaving,  because of invalidness of NUMA hint faults. Depending on temporal and spatial distribution of hot pages, the dynamic page migration can degrade  performance compared to no migration.
%Depending on the hot page distribution in time and address space, the dynamic page migration can lose application performance, compared with no migration. 
We also observe that the old-fashioned NUMA first touch and Tiering-0.8~\cite{tiering0.8} (the most recent Linux Patch for AutoNUMA to support memory tiering) is very effective, outperforming a set of page migration and interleaving solutions.

%%%\textbf{Existing page management systems for tiered memory face challenges on the CXL memory.} 
 
\section{Background}
\label{sec:background}
%In this section, we introduce background information.
%\textcolor{red}{(Jie Liu will work on it.)}

\subsection{Compute Express Link}
The CXL specification defines three protocols: \texttt{CXL.io}, \texttt{CXL.cache}, and \texttt{CXL.mem}. There are three types of CXL devices. The type-3 device is related to our evaluation. Such a device supports \texttt{CXL.io} and \texttt{CXL.mem}, and is used for memory bandwidth or capacity expansion in memory tiering. %%and storage class memory in memory tiering.
%%%for initial discovery and configuration of CXL devices, \texttt{CXL.cache} for CXL-devices to access the host memory using PCIe protocol features, and \texttt{CXL.mem} for the host access to CXL device memory. The Home Agent (HA) on the CPU and CXL controller on the device manage the interaction between the host and CXL device. The HA administers the \texttt{CXL.mem} protocol and seamlessly presents the CXL memory to the host as memory on a remote NUMA node.
%\textcolor{check}{The CXL specification defines three separate protocols: \texttt{CXL.io}, \texttt{CXL.cache}, and \texttt{CXL.mem}. \texttt{CXL.io} is responsible for the initial discovery and configuration of CXL devices. \texttt{CXL.cache} enables CXL-devices to access the host (CPU) memory using the PCIe protocol features. \texttt{CXL.mem} is dedicated to memory accesses from the host to the CXL device. The interaction between the host and CXL device is managed by the Home Agent (HA) and the CXL controller, located on the CPU and the device, respectively. seamlessly presents the CXL memory to the host as if it were memory on a remote NUMA node.}
%\textcolor{check}{\texttt{CXL.mem} provides three coherence models for CXL exposed host-managed device memory (HDM): HDM-H (host-only coherent and used  for type-3), HDM-D (device coherent relying on \texttt{CXL.cache} to manage coherence with the host and used for type-2), and HDM-DB (device coherent using back-invalidation and used for type-2 or type-3).}
The CXL specification has been going through three major versions: 1.1, 2.0, and 3.0. \textcolor{revision}{CXL} 1.1 focuses on directly-attached CXL devices, \textcolor{revision}{CXL} 2.0 incorporates switch-based pooling, and \textcolor{revision}{CXL} 3.0 supports switch-less pooling and higher bandwidth.

Most of the real CXL devices nowadays are host-managed device memory with host-only coherent (HDM-H) \textit{using CXL 1.1}. The three devices for our evaluation are among them.

%CXL.mem enables CPUs and other CXL devices to access device memory as cacheable memory. CXL.mem makes it possible for memory attached to a device to be cacheable (referred to as 'Host managed Device Memory' – HDM), similar to the host memory, resulting in a host’s uniform view across HDM and host memory.

\begin{comment}
\begin{figure}[tb!]
	\centering
	\includegraphics[width=1.0\linewidth]{./figures/machine_org.pdf}
	\caption{Memory organization of our evaluation platforms.  \textcolor{red}{(delete it?)}}
	\centering
	\label{fig:machine_org} 
\end{figure}
\end{comment}

\begin{table}[t]
\caption{Three systems with CXL devices.}
\vspace{-20pt}
\label{tab:system_config}
\begin{center}
\resizebox{0.49\textwidth}{!}{
\begin{tabular}{|c|c|c|}
\hline
{\textbf{Sys}} & {\textbf{Component}} & {\textbf{Description}} \\
\hline \hline
\multirow{8}{*}{\textbf{A}} & OS (kernel)     & Ubuntu 22.04 LTS (Linux kernel v6.2.15) \\
 & \multirow{2}{*}{CPUs}        & $2\times$ AMD EPYC 9354 CPUs @3.8 GHz, \\ 
 & & 32 cores and 512 MB LLC per CPU \\
 & PCIe     & PCIe 5.0, speed 32GT/s,  16 lanes \\
& & \\
 &\multirow{2}{*}{ Memory}         &  Socket 0: $12\times$ DDR5-4800 channels, memory 768GB \\\
 & & Socket 1: $12\times$ DDR5-4800 channels, memory 768GB \\
 & & \textcolor{revision}{max bandwidth 460.8 GB/s per socket}\\
  & & \\
 & \multirow{2}{*}{CXL A}       & Single channel DDR5-4800, memory  \\
 & & 128 GB, max bandwidth 38.4 GB/s per channel \\
\hline \hline
\multirow{8}{*}{\textbf{B}} & OS (kernel)     & Fedora Linux 36 (Linux kernel v6.6.0-rc5) \\
 & \multirow{2}{*}{CPUs}        & $2\times$ Intel(R) Xeon(R) Platinum 8470 CPU @2.0GHz, \\ 
 & & 52 cores and 210 MB LLC per CPU (Saphire Rapids) \\
  & PCIe     & PCIe 5.0, speed 32GT/s,  16 lanes \\
 & & \\
 &\multirow{2}{*}{ Memory}         &  Socket 0: $8\times$ DDR5-4800 channels, memory 1TB \\\
 & & Socket 1: $8\times$ DDR5-4800 channels, memory 1TB \\
  & & \textcolor{revision}{max bandwidth 307.2 GB/s per socket}\\
  & & \\
 & \multirow{2}{*}{CXL B}       & Single channel DDR5\textcolor{revision}{-8000}, memory  \\
 & & 64 GB, max bandwidth 64.0 GB/s per channel \\
\hline \hline
\multirow{8}{*}{\textbf{C}} & OS (kernel)     & Ubuntu 22.04 (Linux kernel v6.2.15) \\
 & \multirow{2}{*}{CPUs}        & $2\times$ Intel(R) Xeon(R) Gold 6438Y+ @2.0GHz, \\ 
 & & 32 cores and 60 MB LLC per CPU  \\
  & PCIe     & PCIe 5.0, speed 32GT/s,  16 lanes \\
  & & \\
  &\multirow{2}{*}{ Memory}         &  Socket 0: $8\times$ DDR5-4800 channels, memory 512GB \\\
 & & Socket 1: $8\times$ DDR5-4800 channels, memory 512GB \\
  & & \textcolor{revision}{max bandwidth 307.2 GB/s per socket}\\
  & & \\
 & \multirow{2}{*}{CXL C}       & Dual channel DDR5-6200, memory  \\
 & & 128 GB, max bandwidth 48.4 GB/s per channel \\
\hline
\end{tabular}
}
\end{center}
\vspace{-18pt}
\end{table}

%SR3:
%OS Kernel: Ubuntu 22.04, Kernel 6.2.15
%CPUS: 2 x Intel(R) Xeon(R) Gold 6438Y+ @ 4 GHz
%Memory: Socket 0: 8 x 64GB DDR5-4800, memory 512 GB
%Memory: Socket 1: 8 x 64GB DDR5-4800, memory 512 GB
%CXL C: Dual Channsl DDR5-489: 128GB

%SR0:
%CXL B: 64GB, single channel,  DDR5 4800

\subsection{CXL Systems for Evaluation}
%\textcolor{red}{(add machine organisations for three CXL machines)}
CXL requires compatible hardware in both the CPUs and peripheral devices. The 4th-generation Intel Xeon Scalable Processors (such as Sapphire Rapids) and the 4th-generation AMD EPYC Processors (such as Genoa) are among the first mainstream server CPUs to support the CXL 1.1. Several CXL memory devices have been developed as commercial products by leading hardware manufacturers such as Micron.  %Montage, Samsung, and Micron. 
%In this paper, we evaluate the performance of three genuine CXL-ready systems using ASIC-based hard CXL IPs, listed in Table~\ref{tab:system_config}. 
We use three CXL devices from three vendors. See Table~\ref{tab:system_config}. %%%summarizes the three systems equipped with the devices. 
%hence covering a spectrum of possible CXL implementations in the market. 

\begin{comment}
%Figure~\ref{fig:machine_org} shows the memory organization of the system A (listed in Table~\ref{tab:system_config}). 
\textcolor{red}{ (simplify) The system A listed in Table~\ref{tab:system_config} has two sockets (0 and 1), and a CXL device is attached to the socket 1 by a CXL link over PCIe 5.0. This system has two AMD EPYC 9354 CPUs. Accessing the CXL memory from CPU 0 has to go through the HyperTransport interconnect, which leads to longer latency than accessing from CPU 1. Different from the traditional NUMA node, the CXL memory device lacks CPU cores and caches, and does not have an expensive interconnect between the CXL IP and the device's memory controller. Hence, the CXL memory is represented as a CPU-less NUMA node in the system A. From the view of a CPU, there are three NUMA nodes: local DDR (or LDRAM), remote DDR on the other socket (RDRAM), and CXL memory. The system B has the same organization as the system A. %%%The Sapphire Rapids processors on the system B need the experimental ``legacy mode'' to be enabled in the BIOS to use the attached CXL memory. 
The system C has a slightly different organization: the CXL device is attached to the socket 0 (instead of socket 1). }
\end{comment}

The system A in Table~\ref{tab:system_config} has two sockets (0 and 1), with a CXL device attached to Socket 1 by CXL link over PCIe 5.0. Accessing CXL memory from Socket 0 goes through HyperTransport interconnect, leading to longer latency than accessing from Socket 1. From the view of a CPU, there are three NUMA nodes: local DDR (LDRAM), remote DDR on the other socket (RDRAM), and CXL memory. The system B has the same organization as A. The system C has a different organization: the CXL device is attached to Socket 0 (not 1).

\section{Basic Performance Characteristics}
\label{sec:basic_perf}

\begin{figure}[tb!]
	\centering
	\includegraphics[width=0.85\linewidth]{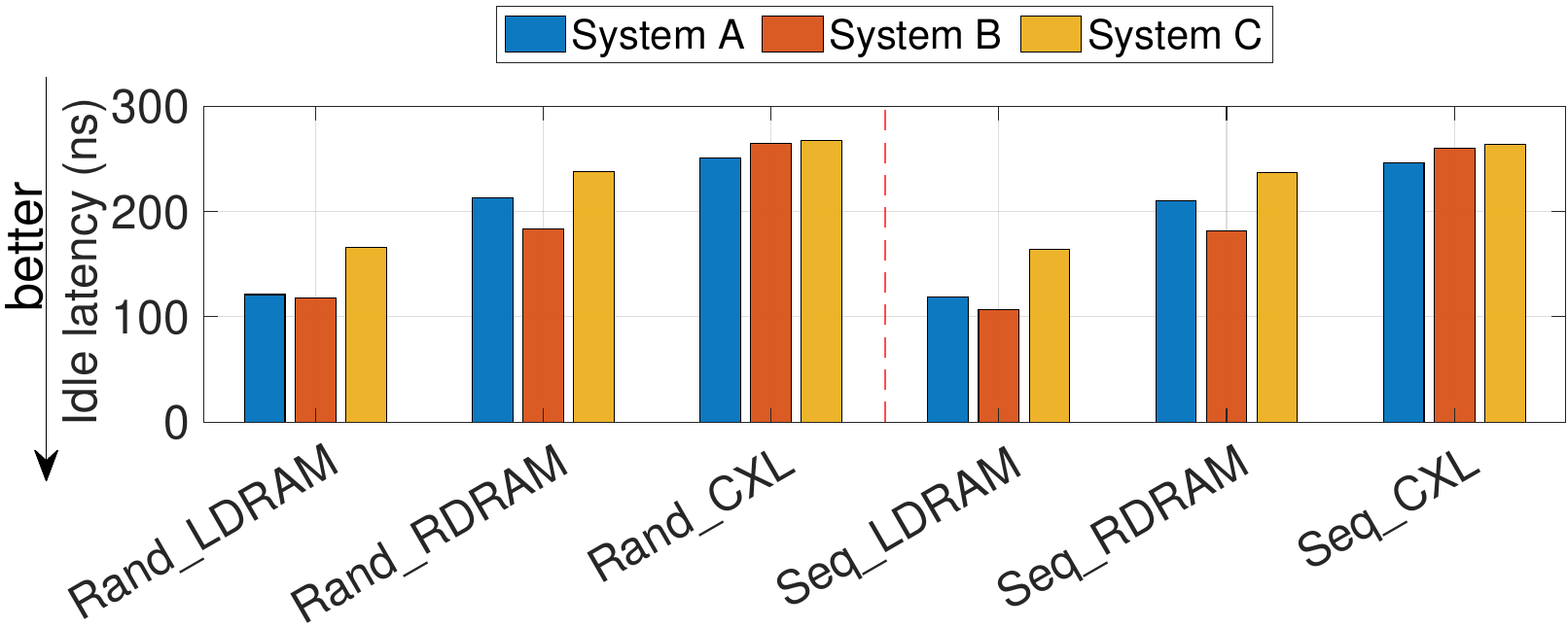}
    \vspace{-10pt}
	\caption{Load latency with random and sequential accesses to a cache block.}
	\centering
	\label{fig:cxl_latency} 
    \vspace{-15pt}
\end{figure}
%The idle latency of the memory devices on different servers.

\textbf{Evaluation methodology.} We evaluate memory latency and bandwidth using Intel Memory Latency Checker (MLC) \cite{mlc}. MLC disables hardware prefetcher for Intel processors (the systems B and C),  %by writing specific values to some Processor-specific Registers (MSRs) that control hardware prefetchers, 
but cannot do so for AMD processors (the system A). For latency tests, MLC uses typical pointer chasing. For each latency test, we repeat the test 5,000 times, and report the average value after excluding outliers (caused by operating system services and random TLB misses).  For bandwidth tests, we use MLC to perform sequential and random memory accesses. The sequential accesses in combination with thread-level parallelism introduces parallel memory accesses, revealing peak memory bandwidth. %We also change the ratio of read to write to measure bandwidth. 
For each bandwidth test, we repeat the test 2,000 times, and report the average value. %In the rest of the paper, local DDR and remote DDR are denoted by  ``LDRAM'' and ``RDRAM'' respectively.

\begin{figure}[tb!]
	\centering	\includegraphics[width=0.9\linewidth]{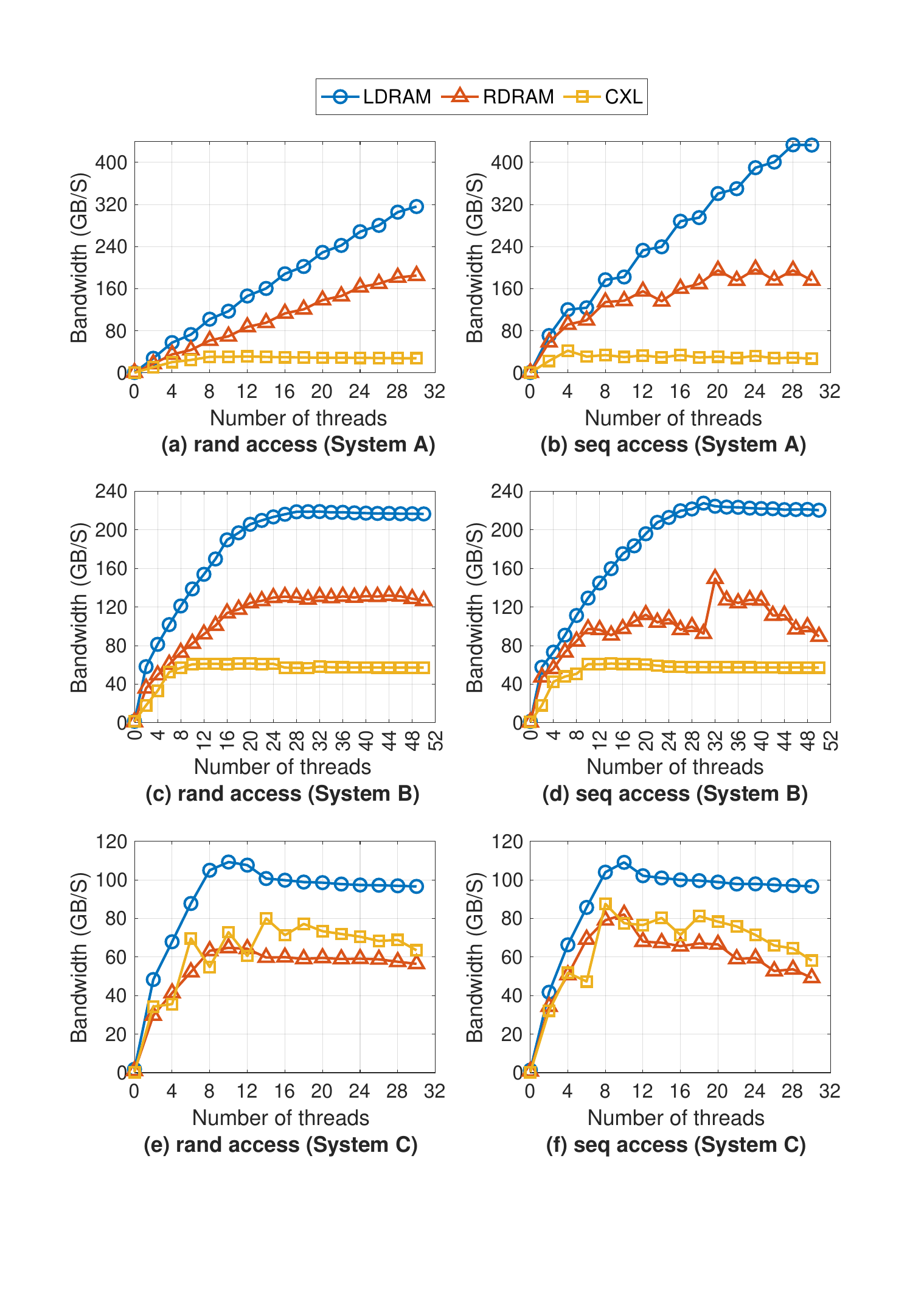}
 \vspace{-10pt}
	\caption{Bandwidth scaling for data loading. \textcolor{revision}{Note that the scales of the figures for the three systems are different.}}
	\centering
	\label{fig:rm_all_bw} 
 \vspace{-15pt}
\end{figure}
% Server A
%random, peak bw. LDRAM: 316.1305 (9.8%), RDRAM: 184.7605 (16.8%), CXL: 31.0249
%sequential, peak bw. LDRAM: 432.9590 (7.8%), RDRAM: 196.8185 (17.1%), CXL: 33.7218 

% Server B
%random, peak bw. LDRAM: 219.0573 (28.02%), RDRAM: 131.9173 (46.6%), CXL: 61.4003
%sequential, peak bw. LDRAM: 227.7505 (27.0%), RDRAM: 149.5031 (41.1%), CXL: 61.4019

% Server C
%random, peak bw. LDRAM: 109.2496 (73.3%), RDRAM: 64.7472, CXL: 80.0366
%sequential, peak bw. LDRAM: 109.0858 (80.3%), RDRAM: 82.0710, CXL: 87.5520

%%How the bandwidth changes for the different NUMA nodes on the three servers. 

%\begin{figure}[tb!]
%	\centering
%	\includegraphics[width=1.0\linewidth]{./figures/rm_bandwidth.pdf}
%	\caption{How the bandwidth changes as the number of cores increases.}
%	\centering
%	\label{fig:ram_bandiwdth} 
%\end{figure}
% A (rand): 121.1, 213.2, 250.5
% B (rand): 118.3, 183.2, 264.9
% C (rand): 165.7, 237.5, 275.5

% A (seq): 118.9, 210.6, 246.0, 
% B (seq): 107.0, 181.5, 260.1
% C (seq): 163.8, 236.8, 274.0

\textbf{Latency results.} See Figure~\ref{fig:cxl_latency}. Compared with LDRAM, CXL is much slower than one-hop-away NUMA node (RDRAM). In fact, assuming that adding a hop of NUMA distance introduces a constant latency in a system, \textit{the CXL memory is comparable to a two-hop-away NUMA node, in terms of access latency}. Figure~\ref{fig:cxl_latency} shows that the CXL memory from different vendors show quite different latency. For example, for sequential accesses, the CXL memory in the system A adds latency by 153 $ns$, while the CXL memory in the system B adds latency by 211 $ns$, \textcolor{check}{compared to LDRAM}. Since the two systems use the same PCIe and DRAM technologies, such a latency difference mainly comes from the difference in the CXL controller and HA on the CPU. 

\begin{figure*}[t!]
	\centering	\includegraphics[width=0.95\linewidth]{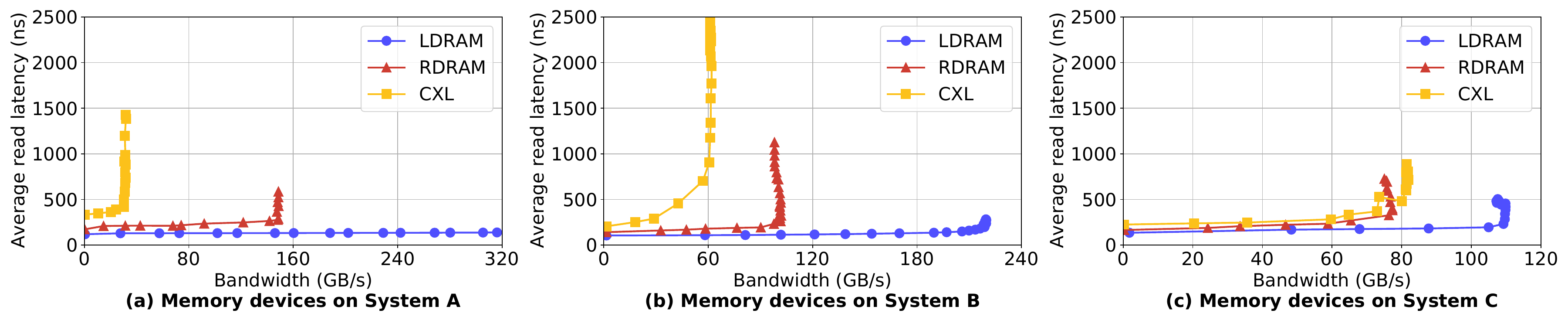}
    \vspace{-12pt}
	\caption{Memory latency and bandwidth under varying load.}
	\centering
	\label{fig:band_latency} 
    \vspace{-17pt}
\end{figure*}

\textbf{Bandwidth results.}  Figure~\ref{fig:rm_all_bw} shows \textcolor{revision}{that bandwidth scaling of LDRAM, RDRAM, and CXL are different as we change the number of threads.}
% Figure~\ref{fig:rm_all_bw} shows the bandwidth scaling as we change the number of threads. %We have three observations.  
%\underline{Basic observation 3.} 
The CXL memory bandwidth is saturated as the number of threads is over 8, while the saturation points for LDRAM and RDRAM are much higher than that in CXL (e.g., 28 and 20 on the system B), %\textcolor{dong}{indicating that the CXL memory bandwidth is limited by single DRAM channel bandwidth.} 
\textcolor{ren}{because the bandwidth of CXL memory is constrained by the bandwidth of a single DDR channel.} 
%\underline{Basic observation 4.} %%%The peak CXL memory bandwidth is much lower than that of LDRAM (the CXL memory bandwidth is 9.8\%, 28.1\%, and 80.3\% of LDRAM bandwidth on the systems A, B, and C respectively).  
The peak CXL memory bandwidth is lower than that of RDRAM (the CXL memory bandwidth is 17.1\% and 46.4\% of the RDRAM bandwidth on the systems A and B respectively), but can be close to the RDRAM bandwidth (see the system C). 

%\underline{Basic observation 5.} 
\textit{The difference in bandwidth scaling between LDRAM, RDRAM, and CXL highlights the importance of distributing memory accesses between them.} For example, in the system B, to maximize the bandwidth usage, we would assign 6, 23, and 23 threads to access CXL, LDRAM, and RDRAM respectively\textcolor{revision}{, because increasing thread counts beyond these points does not improve the bandwidth, shown in Figure~\ref{fig:rm_all_bw}(d)}. \textit{Using the above thread counts can lead to a peak bandwidth of 420 GB/s, larger than any other thread assignment.}  

\begin{comment}
\textcolor{red}{(delete this?)}
\underline{Basic observation 6.} The hardware prefetcher is not very effective for CXL, but effective for LDRAM and RDRAM. See the system A. The system A is an AMD machine where the prefetcher is not disabled by Intel MLC. The other two systems are based on Intel processors and disables the prefetcher. As a result, we do not see performance difference between random and sequence accesses in the systems B and C (sequence accesses should benefit from the prefetcher if turned on and hence outperform random accesses). In contrast, in the system A, LDRAM and RDRAM get performance benefits from the prefetcher: the memory bandwidth is higher. However, we do not see such a performance benefit in the CXL: there is no performance difference between random and sequence accesses. We suspect that slow memory accesses in CXL makes the prefetching mechanism less effective.
\end{comment}

\textbf{Performance under load.} We study memory latency and bandwidth under varying load. Figure~\ref{fig:band_latency} presents the results of how latency and bandwidth vary by gradually increasing the load on memory. For this test, we employ Intel MLC, utilizing 32 threads. In our methodology, each thread performs memory accesses to cache lines and delays for a time interval between two accesses. We vary the time interval from 0 to $80 \mu s$. This setup ensures that each worker thread is engaged in repeated, sequential memory accesses, allowing for a detailed analysis of how varying load conditions impact memory performance. When the time interval is 0  (corresponding to the right side of each subfigure in Figure~\ref{fig:band_latency}), the bandwidth is close to the maximum bandwidth and the latency skyrockets as the queuing effects in hardware dominate. When the time interval is high enough ($80 \mu s$, corresponding to the left side of each subfigure in Figure~\ref{fig:band_latency}), the latency is close to the raw, unloaded latency. 

%%%%\underline{Basic observation 7.} There is a ``knee'' point of latency and bandwidth for each device as we increase the memory access bandwidth. Such a ``knee'' point shows at which the memory device is able to maintain steady bandwidth without suffering from queuing effects in hardware. For example, in Figure~\ref{fig:band_latency}(b), the ``knee'' point for CXL memory occurs when the bandwidth is approaching 32GB, while the ``knee'' point for LDRAM and RDRAM occurs at the bandwidth value approaching to 128GB and 225GB, respectively. 

%%%\underline{Basic observation 8.} The maximum bandwidth of each memory device determines when the ``knee'' point happens. As Figure~\ref{fig:band_latency} shows, the ``knee'' of CXL memory occurs earlier than LDRAM and RDRAM, except for RDRAM on System C. In particular, for System C, the ``knee'' of CXL memory happens when the bandwidth is near to 82GB, and its corresponding maximum bandwidth is 96GB. While for LDRAM and RDRAM, their ``knee'' occurs when the  bandwidth is near 84GB and 110GB respectively. The maximum bandwidth when accessing RDRAM is bounded by the max interconnect bandwidth cross NUMA node, which is 96GB. And the maximum bandwidth of LDRAM is bounded by the DDR version and the number of channels, which is 128GB.

\underline{Basic observation.} The latency of accessing LDRAM and RDRAM can be similar to the CXL memory when reaching their peak memory bandwidth. %Figure~\ref{fig:band_latency} shows that  once the bandwidth consumption of LDRAM and RDRAM is near  the maximum, their latency significantly increases. 
For example, in Figure~\ref{fig:band_latency}(c), once the bandwidth consumption of LDRAM and RDRAM approaches the peak (110GB/s for LDRAM and 84GB/s for RDRAM), their latencies are as high as 543 $ns$ and 600 $ns$ respectively, which are pretty close to the CXL memory latency (400 $ns$ - 550 $ns$) when approaching the peak CXL bandwidth under heavy load. \textit{This demonstrates the potential of using CXL as LDRAM and RDRAM under heavy load.}

\vspace{-6pt}
\begin{tcolorbox}
\footnotesize
\textbf{Takeaway}: %The CXL memory performs as a NUMA node with similar latency and much lower bandwidth (or comparable bandwidth), compared to RDRAM. 
CXL memory performs as a NUMA node with latency similar to RDRAM but lower (or comparable) bandwidth. Different from the traditional NUMA node, the CXL memory is unique in terms of performance scalability and performance under heavy load. 
\end{tcolorbox}

\section{CXL for Large Language Models}
\label{sec:llm}

LLMs are crucial for powering various AI applications, but their substantial memory footprint presents deployment challenges. We study the performance of LLMs with \textit{tensor offloading} techniques using the CXL memory, a promising solution to address the constraint of GPU memory capacity. Tensor offloading moves tensors out of GPU memory when they are not in use, allowing for the execution of larger models that exceed the GPU's memory capacity. \textit{The evaluation is conducted on the system A}, featuring an NVIDIA A10 GPU with 24GB memory, connected to the host CPU via PCIe Gen 4, offering a maximum bandwidth of 32GB/s.

The CXL memory on the system A uses CXL 1.1, which does not allow the GPU  to directly access the CXL memory; instead, the accesses must go through the CPU. Under CXL 1.1, the data path from the GPU to the CXL memory is ``GPU - PCIe - CPU - PCIe - CXL memory'', longer than the direct ``CPU - PCIe - CXL memory'' path. The data path in CXL 1.1 is different from the CXL devices peer-to-peer access supported in CXL 3.1 (using  \textcolor{check}{``GPU - PCIe - CXL memory''}). We explore the effects of the CXL 1.1 data path on the data transfer bandwidth and latency.

\begin{figure}[!tbp]
  \centering
   \includegraphics[width=0.8\columnwidth]{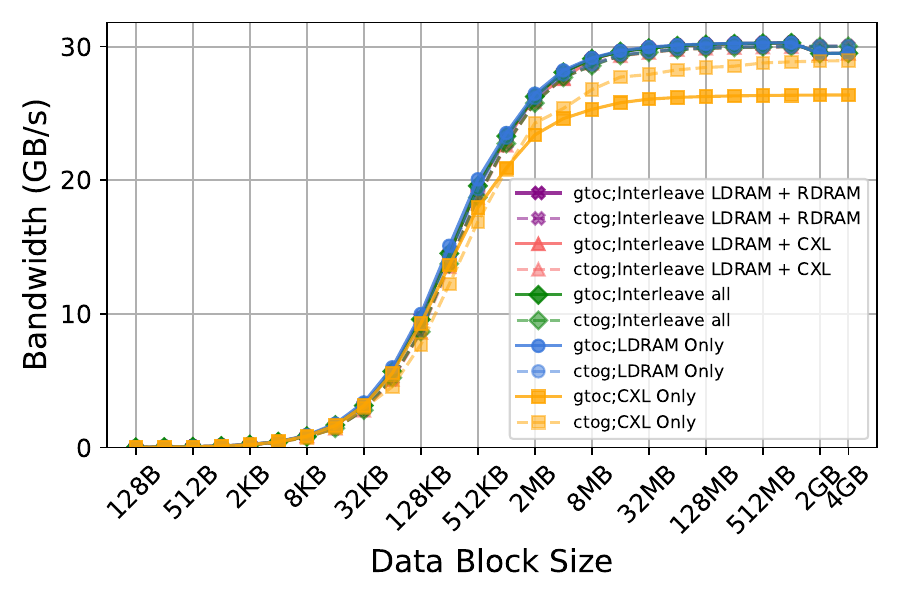}
\vspace{-15pt}
  \caption{Data transfer bandwidth between GPU  (i.e., ``g'') and CPU (i.e., ``c''). }
  % \textcolor{red}{((1)Make the words in the figure larger, as large as the words in paper or slightly bigger. (2) Use the template same as Figure 4 and 5 to draw the figure to make the style the same)
   \vspace{-15pt}
\label{fig:asy-bw}
  % \vspace{-5pt}
\end{figure}

\textbf{Bandwidth results.} %With the method 1, the CXL memory and DDR form a ``single'' memory node using memory interleave. 
We measure memory access bandwidth by repeatedly and randomly copying data blocks between the GPU memory (labeled as ``g'') and CPU memory (labeled as ``c'') using various page allocation strategies. The size of the data blocks varies from 128 byte to 4GB, illustrated in Figure~\ref{fig:asy-bw}. We use membind~\cite{numactl1} to specify the memory device as the source or destination for data transfers.

%We also use membind in numactl~\cite{numactl1} such that data block transfer can happen between GPU and a specific memory node (DDR or CXL memory). We have the following observations.

\begin{comment}
    
\begin{itemize}[leftmargin=*,noitemsep,topsep=0pt]

\item Various interleaving polices leads to very similar peak memory bandwidth. For example, comparing ``Interleave ALL'' and ``Interleave LDRAM + RDRAM'', the peak bandwidth difference is less than 3\%, and using CXL memory does not increase bandwidth, because the PCIe interconnect between CPU and GPU becomes a performance bottleneck. 

\item Binding to the CXL memory only, the peak bandwidth is about 25 GB/s, 17\% smaller than the peak CXL bandwidth (\~30 GB/s) we measure on CPU (see Figure~\ref{fig:rm_all_bw}). 
\end{itemize}
\end{comment}

\underline{LLM basic observation 1. } The memory bandwidth for the GPU access to the memory hierarchy with CXL is constrained by the PCIe bandwidth between the CPU and GPU, due to the absence of peer-to-peer access support in CXL 1.1. 

\textcolor{check}{Counter-intuitively, using the CXL memory does not increase the memory bandwidth for the GPU. As shown in Figure~\ref{fig:asy-bw}, the peak memory bandwidth is similar across various memory interleaving policies (the difference is less than 3\%).}
%configurations with variations in memory allocation strategies less than 3\%. 
This lack of bandwidth increase with the CXL memory is primarily due to the PCIe interconnect between the CPU and GPU acting as a performance bottleneck.
%Counterintuitively, using CXL memory does not increase bandwidth when GPU access memory. The memory bandwidth, while accessing various memory devices under different interleaving policies, results in a peak memory bandwidth that is  similar across configurations, as shown in figure~\ref{fig:asy-bw}. The variations in memory allocation strategies in both directions are less than 3\%.

%Memory with CXL 1.0/1.1 protocol can not be used as bandwidth expansion for GPU memory. Due to lack of peer-to-peer access between GPU and CXL memory, the GPU access CXL memory bandwidth is bounded by PCIe that connect GPU to CPU. 
%%Therefore there is no difference in bandwidth between different memory allocation strategies the both direction. 

%\underline{In conclusion}, the longer data path has negative impacts on CXL memory bandwidth. Using the interleaving policy, although increasing memory bandwidth (compared to using a single NUMA node alone), the performance is  bottlenecked by the interconnect between CPU and GPU when using CXL 1.0.

%%Memory with CXL 1.0/1.1 protocol can not be used as bandwidth expansion for GPU memory. Due to lack of peer-to-peer access between GPU and CXL memory, the GPU access CXL memory bandwidth is bounded by PCIe that connect GPU to CPU. 
%%Therefore there is no difference in bandwidth between different memory allocation strategies the both direction. 

\begin{figure}[!t]
  \centering
   \includegraphics[width=0.8\columnwidth]{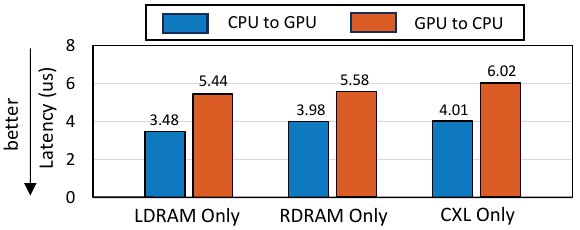}
    \vspace{-10pt}
  % \caption{Data transfer latency between the GPU and CPU. The data size is 64 bytes.}
    \caption{64-byte data transfer latency between the GPU and CPU.}

   %\vspace{-5pt}
\label{fig:ai-latency}
 \vspace{-20pt}
\end{figure}

%\textbf{Tensor Bandwidth Behavior.} In order to profile the bandwidth behavior, we execute bandwidth profiling script from FlexGen \cite{flexgen} where we execute data movement function 3 times as warm-up, then we execute the function 20 times as one round. In total, we have 5 rounds and we get the average value as final bandwidth. As shown in Figure \ref{fig:asy-bw}, we use numactl \cite{numactl} interleave to control data access, increase the data movement size and exchange the data path like ctog (CPU to GPU) and gtoc (GPU to CPU). We observe that bandwidth asymmetry behavior exists when transferring large data (8 GB) from CPU to GPU and from GPU to CPU.

%We use a microbenchmark \cite{CUDAMicroBench} to test the latency of host-device memory copy and we fix data size to 64 bytes (cache line size). We execute the memory-copy operations from device (labeled as 'd') to host (labeled as 'h') and host to device for 100,000 times separately and get the average value as final latency. As shown in Figure \ref{fig:ai-latency}, we have the following observations.

\textbf{Latency results.} We develop a microbenchmark to 
measure the data transfer latency. The microbenchmark runs on CPU 1 (the CPU close to the CXL memory) and uses \texttt{cudaMemcpy()} for data transfer. The benchmark repeatedly transfers a 64-byte data (a cache block) between the CPU and GPU. The transfer happens 100K times, and we report the average time for one transfer. We use membind, similar to the bandwidth tests. Figure~\ref{fig:ai-latency} shows the results.
%On CPU, we evaluate various interleaving policies (e.g., interleaving 0 and 1, and interleaving 1 and 2). \textcolor{red}{For an interleaving policy using multiple memory components, a data block on CPU is randomly accessed: it can be located on any component. As a result, the latency for an interleaving policy is the average memory access latency of multiple memory components.} 
%Figure~\ref{fig:ai-latency} shows the results. We have the following observations.

\begin{comment}

\begin{itemize}[leftmargin=*,noitemsep,topsep=0pt]
\item The data transfer latency between GPU and CPU (including the accesses to the CXL memory) is at least $10 \times $ larger than the memory access latency on CPU. For example, on CPU, the memory access latency on CXL is about 250 $ns$ (see Figure~\ref{fig:cxl_latency}). However, the data transfer latency between GPU and CPU is about 3-6 $\mu s$. Such a large increase in latency is longer than extra PICe latency (about 150 $ns$) in the data path. The CUDA runtime overhead (including the overhead in GPU driver) may be the reason. 

\item Data transfer latency to the CXL memory is longer than the transfer latency to other memory nodes. For example, from GPU to CPU, the data transfer time is 5.44 $\mu s$, 5.58 $\mu s$ and 6.02 $\mu s$ for LDRAM Only, RDRAM Only and CXL Only respectively.%the memory nodes 0, 1 and 2 (the CXL memory) respectively. 
\end{itemize}
    
\end{comment}

\underline{LLM basic observation 2.} Accessing the CXL memory from the GPU can result in longer latency than expected.

The difference of data transfer latency between ``GPU - CXL memory'' and ``GPU - CPU memory'' (in Figure~\ref{fig:ai-latency}) is greater than the latency difference between ``CPU - CXL memory'' and ``CPU - CPU memory'' (in Figure~\ref{fig:cxl_latency}). For example, accessing the CXL memory from the GPU is, on average, 500 $ns$ longer than accessing the CPU memory from the GPU. In contrast, accessing the CXL memory from the CPU is only 120 $ns$ longer than accessing the CPU memory from the CPU. The longer latency-difference from the GPU side comes from  the longer data-path between the GPU and CXL memory.  %caused by the extra PCIe latency.
%Accessing CXL memory from GPU is 500 $ns$ (on average) longer than accessing DRAM on host CPU, as shown in Figure~\ref{fig:ai-latency}. Meanwhile, accessing CXL memory from the CPU is 120 $ns$ longer than accessing DRAM on the host CPU. The longer latency from the GPU is due to the extended data path caused by the extra PCIe latency in GPU accesses.

We study the implication of using the CXL memory for LLM \textcolor{revision}{ training and inference} using tensor offloading \textcolor{revision}{on System A}, as follows.

\begin{comment}
\begin{itemize}[leftmargin=*,noitemsep,topsep=0pt]
\item Asymmetry exists in data transfer between host and device where device to host will consume more time because of longer data transfer path shown in Figure \ref{fig:machine_org}. When transfer data from device to host, including CXL memory will add more latency overhead.

\item When transfer data from host to device, only using local DDR (membind 1) will give the least latency while local DDR + remote DDR (interleave 0,1) gives the largest latency.
\end{itemize}
\end{comment}

\begin{comment}
\begin{table}[ht]
\centering
\caption{Evaluation configuration for FlexGen.}
\label{tab:flexgen-system-configuration}
\begin{tabular}{|l|l|}
\hline
\textbf{CPU}          & AMD EPYC 9354 32-Core Processor \\ \hline
\textbf{GPU}          & 24GB NVIDIA A10 Tensor Core GPU \\ \hline
\textbf{Main memory}  & 196GB DDR5 for each node           \\ \hline
\textbf{NVMe memory}  & 128GB          \\ \hline
\textbf{CXL memory}  & 128GB          \\ \hline
\end{tabular}
\end{table}
\end{comment}

\subsection{LLM Training}

\subsubsection{ZeRO-Offload Background}
%\begin{comment}
    
\begin{figure}[!tbp]
  \centering
   \includegraphics[width=.85\columnwidth]{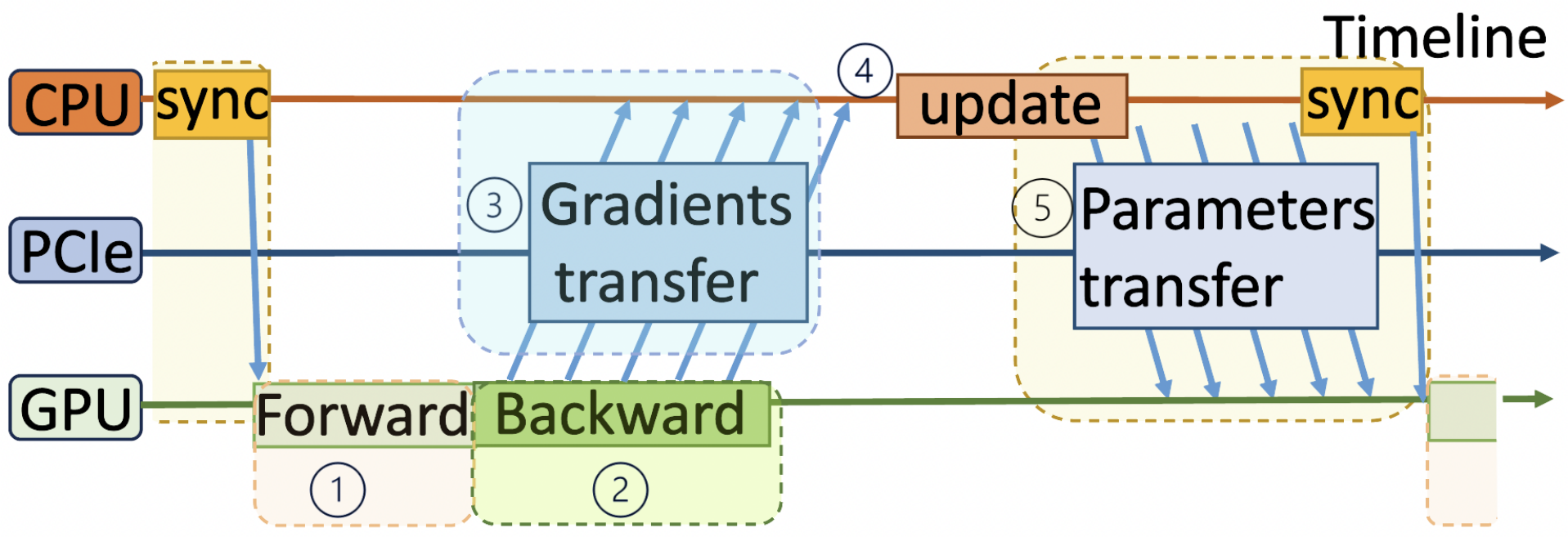}
     \vspace{-10pt}
  \caption{Overview of ZeRO-Offload in a training step.}
   %\vspace{-5pt}
\label{fig:zero-offload}
  \vspace{-18pt}
\end{figure} 
%\end{comment}

\begin{comment}
    
  \begin{table}[!tbp]
\centering
\caption{BERT configurations for  evaluation}
% \vspace{-10pt}
\begin{tabular}{|c|c|c|c|c|}
\hline
\# params & batch size &  \# layer & hidden  & \# att heads  \\ 
 \hline
\hline
110 million& 32  & 12 & 768 & 12\\ \hline
336 million & 14  & 24 & 1024 & 16\\ \hline
4 billion & 2  & 48 & 2560 & 40\\ \hline
\end{tabular}
\label{tab:bert_model_config}
\end{table}

% \vspace{-5pt}

\begin{table}[!tbp]
\centering
\caption{GPT2 configurations for evaluation}
\begin{tabular}{|c|c|c|c|c|}
\hline
\# params & batch size &  \# layer & hidden  & \# att heads  \\ 
 \hline
\hline
4 billion & 13  & 64 & 2304 & 32\\ \hline
6 billion & 8  & 53 & 3072 & 32\\ \hline
8 billion & 3  & 72 & 3072 & 32\\ \hline
\end{tabular}
\label{tab:gpt2_model_config}
\end{table}
\end{comment}

ZeRO-Offload~\cite{atc21:zerooffload} is commonly employed in the industry to enable larger LLM \textit{training} using smaller GPU memory. To save the GPU memory, ZeRO-Offload efficiently offloads full-precision model parameters, gradients, and optimizer states (e.g., momentum and variance) to the CPU memory, moving them back to GPU memory when needed. Figure~\ref{fig:zero-offload} depicts the workflow of ZeRO-Offload.
Specifically, it \circled{1} performs forward and \circled{2} backward computation on GPU; and \circled{3} offloads gradients to the CPU memory during the backward step. To reduce the overhead of tensor movement between CPU and GPU \circled{4}, ZeRO-Offload performs optimization computation (e.g., the ADAM optimizer) on the CPU; and \circled{5} moves updated parameters from the CPU memory to the GPU memory before the next forward step. This strategy minimizes data movement volume between the GPU and CPU memory for each training step. As a result, ZeRO-Offload enables 10$\times$ larger model training on a single GPU with 1.4$\times$ higher throughput.

\begin{figure}[t!]
  \centering
   \includegraphics[ width=0.9\columnwidth]{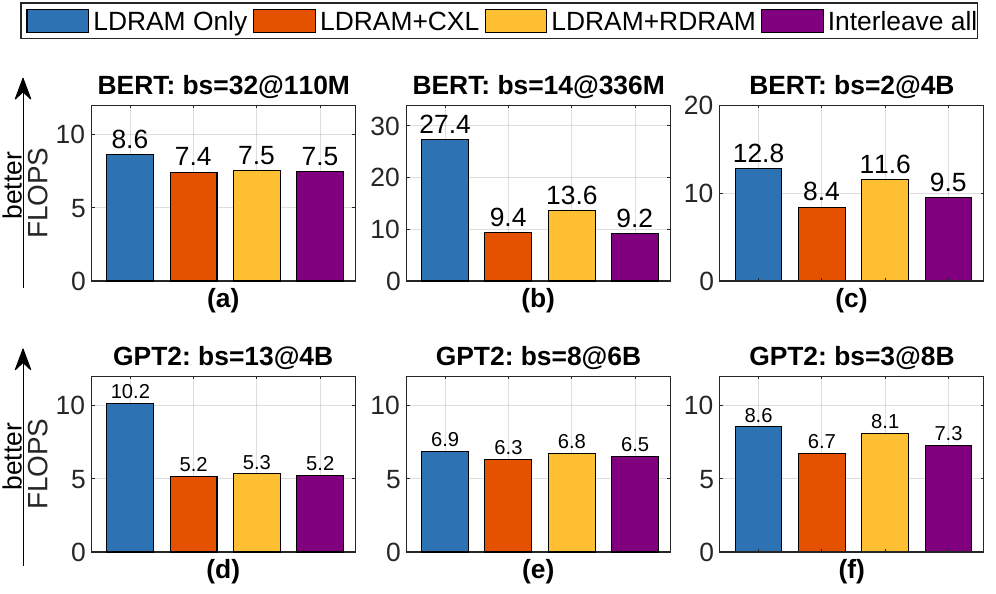}
   \vspace{-10pt}
  \caption{Performance with various interleaving policies and model sizes  for BERT and GPT2. }
   %\vspace{-5pt}
\label{fig:zero-offload_gpt2}
  \vspace{-10pt}
\end{figure} 

\subsubsection{Using CXL Memory}
\begin{figure}[tb!]
  \centering
   \includegraphics[ width=0.95\columnwidth]{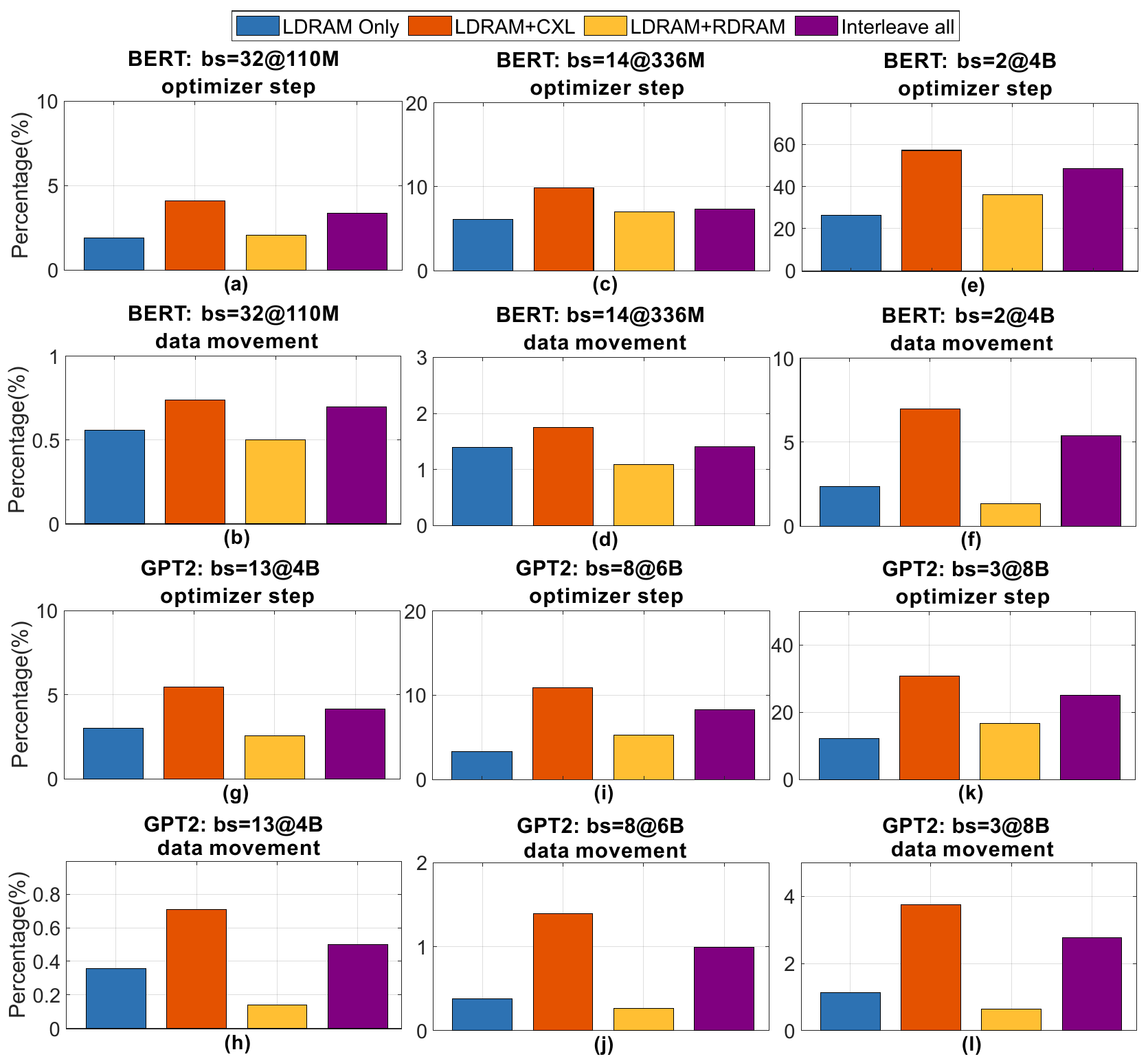}
\vspace{-10pt}
  \caption{Performance of the optimizer and data movement under various interleaving. The percentage numbers are in terms of total training time.}
\label{fig:zero-offload_bert_gpt2_percentage}
  \vspace{-15pt}
\end{figure}

We evaluate two LLMs: BERT \cite{devlin2018bert} and GPT2 ~\cite{radford2019language}. For BERT, we consider three configurations: 110 million (base), 340 million (medium), and 4 billion (large) parameters. For GPT2, we evaluate models with 4, 6, and 8 billion parameters. Figure~\ref{fig:zero-offload_gpt2} shows the performance with various interleaving policies and model sizes. We use the notation ``bs=effective batch size@model size'' to represent the batch size and number of model parameters. \textcolor{revision}{For a given model size, the batch size is chosen to be the maximum without causing an out-of-memory (OOM) error on the GPU.} To better understand the performance, we break it down to the ``optimization step'' (i.e., the ADAM optimizer on the CPU, which is exposed to the critical path) and ``data movement'' (i.e., the gradient transfer from the GPU to the CPU and the parameter transfer from the CPU to the GPU). \textcolor{check}{The memory capacities for various interleaving policies (i.e., LDRAM only, LDRAM+CXL, LDRAM+RDRAM, and ``interleave all'') are 196GB, 324GB, 392GB, and 520GB.} Figure~\ref{fig:zero-offload_bert_gpt2_percentage} shows the results, including data movement exposed to the critical path.  %%respectively.}

%\color{red}{delete? The numbers on top of each bar are the execution time in seconds.}

\textbf{Evaluation results.} \underline{LLM training observation 1.} Using the CXL memory brings little performance improvement or even negative performance impact to ZeRO-Offloading. 

%When the model size is small, the optimizer step takes a large portion of the overall training time and using the CXL memory slows down the training performance. The training does not get bandwidth benefit of using CXL.  

%With O(100M) and up to 4B parameters, BERT is a small model. Figure~\ref{fig:zero-offload_bert_multi_model_flops} shows that using DDR0+CXL and 2DDR+CXL, BERT consistently performs worse than using DDR alone. For example, with ``bs=14@336M'', the best CXL performance (DDR+CXL) is $2.9 \times$ worse than DDR only. 

See GPT2. Figure~\ref{fig:zero-offload_gpt2} shows that the performance difference between LDRAM+RDRAM, LDRAM+CXL, and ``interleave all'' is less than 5\% for the 4B and 6B models. For the 8B, LDRAM outperforms `interleave all'' by 14\%, and LDRAM+RDRAM outperforms LDRAM+CXL by 16\%. There is no performance benefit of using CXL. %larger memory bandwidth. 

\begin{comment}
Figure~\ref{fig:zero-offload_bert_gpt2_percentage} reveals reasons for above observations. 
In GPT2, data movement accounts for a small portion of the training time (less than 5\% in all cases). Due to the bottleneck in the extra PCIe, data movement does not benefit from the higher bandwidth offered by interleaving CXL and DDR. Instead, it suffers from the longer access latency of CXL memory. As a result, compared to using LDRAM, using CXL increases data transfer time and has a minor impact on the overall training time.
The optimizer, which runs on the CPU and is sensitive to memory latency, takes a larger portion of the training time compared to data movement. Using CXL significantly increases the optimizer's execution time by 4\% to 22\% compared to using LDRAM. When the optimizer accounts for a very large portion of the training time (e.g., 31\% when bs=3@8B), the slowdown in the optimizer causes a substantial decrease in training throughput.
\textcolor{red}{For example, when bs=3@8B, using "Interleave all" performs 11\% worse than LDRAM+RDRAM due to an 8\% performance degradation in the optimizer. }
Similar observations are made for BERT.
\end{comment}   

Figure~\ref{fig:zero-offload_bert_gpt2_percentage} reveals three reasons for the above observation. 
% \textcolor{red}{overlap in performance breakdown results?}
\textit{(1)} The data movement takes a rather small portion of the training time in GPT2 (less than 5\% in all cases). %%The data transfer volume in each data transfer is also small \textcolor{red}{(how small?)}, making the data transfer sensitive to the memory latency. 
\textit{(2)} Because of the bottleneck in the extra PCIe, the data movement does not get benefits from the better bandwidth offered by interleaving CXL and DDR (the CPU memory). Instead, the data movement suffers from longer access latency of the CXL memory. As a result, compared with using LDRAM, using CXL actually increases data transfer time and has a minor impact on the training time. 
\textit{(3)} The optimizer takes a larger portion of the training time, compared to the data movement. The optimizer happens on the CPU and is sensitive to memory latency. Using CXL increases its execution time (2\%-18\%), compared to using LDRAM. 
%When the optimizer takes a very large portion of the training time 
When the batch size is small, the optimizer takes a significant portion of the training time (e.g., 31\% when bs=3@8B). In such cases, optimizer slowdown substantially decreases overall training throughput.  For example, when bs=3@8B, using ``interleave all'' performs worse than LDRAM+RDRAM by 11\% due to the worse performance in the optimizer by 8\%. We have the similar observations for BERT. 

% GPT2 bs=3@8B optimizer time percentage
% 8.68, 30.788, 16.693, 24.97
% 24.97-16.69=8.21

% GPT2 percentage [optimizer, datamovement]
% percentage_4B = [1.552,  0.1832;
%          5.629, 0.7287;
%          2.535, 0.1465;
%          4.136, 0.5098;
%          ];

% percentage_6B = [3.18,  0.3637;
%          11.286, 1.4451;
%          5.194, 0.2617;
%          8.252, 0.9942;
%          ];
% percentage_8B = [8.68,  0.8183;
%          30.788, 3.7526;
%          16.693, 0.6366;
%          24.976, 2.7647;
%          ];
We expect that using a larger model (such as GPT-3 with 175 billion parameters)\textcolor{revision}{, which has larger numbers of parameters and gradients,} would result in data movement taking a larger portion of training time \textcolor{revision}{to transfer parameters and gradients}. Unfortunately, we cannot evaluate such a large model due to the limited memory capacity on our GPU. 
%We expect that using larger models (such as OPT-175B), the data movement will take a larger portion of the training time. 
In addition, after reducing the data path between the GPU and CXL memory, the CXL memory can play a bigger role in reducing training time.  Also, using the interleaving policy for the optimizer is not good for performance because of the latency sensitive nature of the optimizer. Using the first touch or ``preferred policy'', the optimizer performance can be improved.

\subsection{LLM Inference}
%We particularly focus on transformer models, which are the most common large AI models. 
%%%%The inferences of LLM often face the challenge on limited GPU memory capacity and bandwidth. Each layer in such a model consists of an attention layer (attention computation) and a multilayer perceptron (MLP) layer. The attention layer in FlexGen can be offloaded to CPU. The attention layer is the key for understanding input context. The MLP layer (on GPU) has a series of neural network layers, processing the output from the attention layer. The combination of attention for contextual awareness and MLP for complex data processing is what makes large language models (LLM) highly effective in tasks like text generation and language understanding. 
\begin{comment}
The inferences of LLM often face the challenge on limited GPU memory capacity and bandwidth. Each transformer layer in an LLM comprises an attention layer and a multilayer perceptron (MLP) layer. Notably, the memory requirements of the attention layer increase \textcolor{red}{quadratically???} with the length of the input prompts. 
In this section, we explore how CXL memory, as an expansion of bandwidth and capacity in the memory hierarchy, introduces new opportunities and challenges for LLM inference.
\end{comment}

\subsubsection{FlexGen Background}
\begin{figure}[!tbp]
  \centering
   \includegraphics[width=1.0\columnwidth]{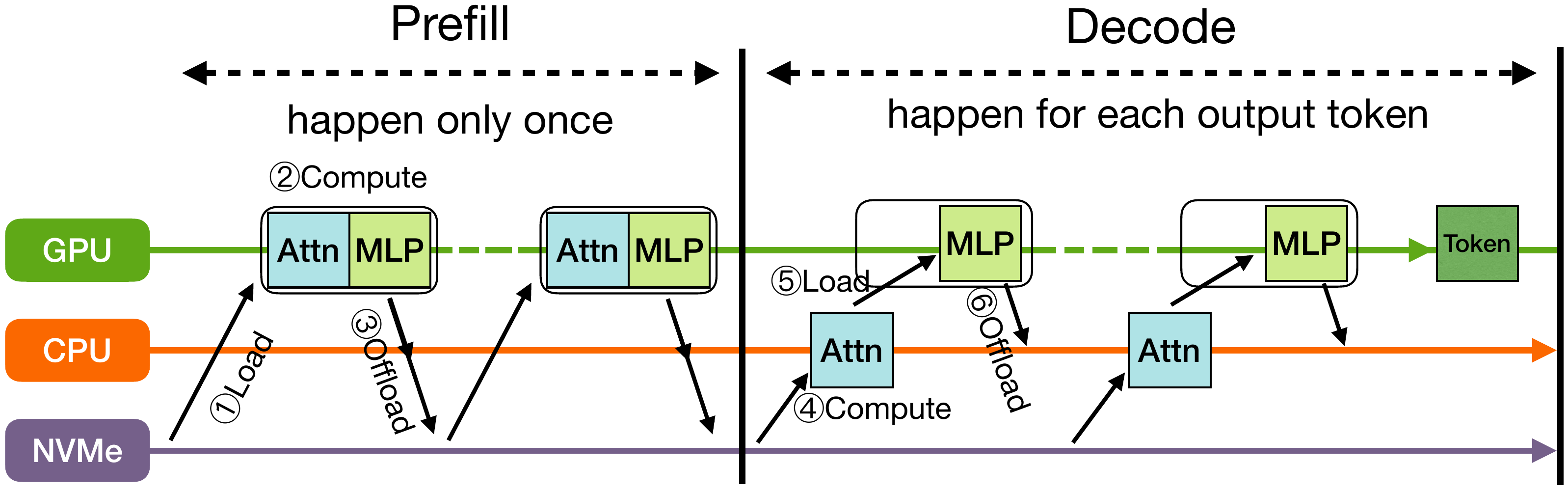}
    \vspace{-20pt}
  \caption{Overview of LLM inference with FlexGen.}
   %\vspace{-5pt}
\label{fig:flexgen}
  \vspace{-5pt}
\end{figure}

%%FlexGen~\cite{flexgen} is a cutting-edge machine learning framework designed for LLM \textit{inference} with constrained GPU memory capacity. 
%%To address memory limitations, FlexGen offloads model parameters, KV cache, and activations to the host CPU memory hierarchy.
%Figure~\ref{fig:flexgen} demonstrates the workflow of FlexGen.
%The inference of LLM generally consists of two stages: prefill and decode. We summarize how tensor offloading happens in the two stages as follows.
LLM inference is memory-consuming. %with the memory requirements of the attention layer increasing linearly with the length of the input prompts.
FlexGen~\cite{flexgen} is a cutting-edge framework designed for LLM \textit{inference} with constrained GPU memory capacity. To address the memory capacity limitation, FlexGen offloads model parameters, KV cache, and activations to the host CPU memory hierarchy \textcolor{check}{(including DRAM and NVMe SSD)}. Figure~\ref{fig:flexgen} shows the workflow of FlexGen. The LLM inference consists of two stages: \textit{prefill} and \textit{decode}.

During the \textit{prefill stage}, which happens only once per inference batch, \circled{1} FlexGen transfers parameters from the CPU memory hierarchy to the GPU. \circled{2} FlexGen executes attention and MLP computation layer by layer on the GPU. \circled{3} At the end of each attention layer, the generated KV cache is offloaded to the CPU memory hierarchy. The \textit{decode stage} generates tokens and significantly influences the overall throughput of the inference process. To minimize tensor movement between the GPU and CPU, \circled{4} FlexGen conducts attention computation directly on the CPU. \circled{5} FlexGen then transfers model parameters and activations generated in the attention layer from the CPU to the GPU for MLP computation, and \circled{6} transfers activations generated in the MLP layer to the CPU for the following computation.

%FlexGen~\cite{flexgen} is a high-throughput generation engine for running large language models (LLM). FlexGen aims to improve LLM inference throughput using limited GPU memory. FlexGen saves GPU memory by offloading tensors to CPU memory and offloading limited computation to CPU. These offloaded tensors are moved back to GPU memory when needed. 
%%%%%utilizes techniques like offloading and decentralized collective inference, which are key in optimizing GPU memory usage and computational efficiency. 

\textbf{Offloading policy and cost model. }  FlexGen allows tensors to be partially placed in the CPU  memory hierarchy and uses a cost model to determine the optimal offloading policy for maximum inference throughput within a memory capacity constraint. The cost model considers latency and bandwidth differences between NVMe and DRAM but does not differentiate among LDRAM, RDRAM, and CXL memory.

\subsubsection{Using CXL Memory}

\textbf{Evaluation setup.} We evaluate LLaMA~\cite{touvron2023llama} with 65 billion parameters and OPT~\cite{zhang2022opt} with 66 billion parameters. \textcolor{revision}{The batch sizes for evaluation are determined by performing a linear programming-based \cite{linearprogramming} policy search to maximize inference throughput while respecting memory and hardware constraints.} The lengths of the input prompts and output tokens are standardized at 2,048 and 256, respectively. 
The evaluation platform has 196 GB LDRAM, 196 GB RDRAM, 128 GB CXL, and 128 GB NVMe. Leveraging GRUB mmap and numactl~\cite{numactl1}, we build the host memory hierarchy with various capacities and media types. 

%We use LLaMA~\cite{touvron2023llama} with 65 billion parameters and OPT~\cite{zhang2022opt} with 66 billion parameters for evaluation. The memory footprint for inference LLaMA and OPT is \textcolor{red }{at least XXGB and XXGB respectively.} The input prompt and output token lengths are 2,048 and 256, respectively. %During the policy search, we eliminate the NVMe SSD but includes the CXL memory. 
%Table~\ref{tab:flexgen-policy-search} shows the configuration used for evaluation.

%We report the performance from the following four observations.
%We use cost model to estimate the throughput under different circumstances for performance comparison. Our evaluations are based on OPT \cite{} and LLaMA \cite{} models, where we set the prompt length as 2048 and the token generation length as 256. GPU batch size, number of GPU batches, and offloading percentage of model weights, KV cache and activation (in Table \ref{tab:flexgen-policy-search}) are given by the cost model for throughput comparison. For different media, we use numactl \cite{} to control tensor access.

% (1) diff in 2 ddr vs ddr + cxl vs ddr + nvme (diff media) (bw specified by policy)
% (2) 1 DDR -> 2 DDR -> 2DDR +cxl, diff bsz -> throughput, bw increase(diff policy) -> perf diff
%           -> 1DDR + cxl

%   (1)(2) runs attention in cpu        
% (3) atten in gpu(pin weights=False) vs attn in cpu(pin weights=False) [2DDR + cxl]

\begin{figure}[!tbp]
  \centering
   \includegraphics[width=1.0\columnwidth]{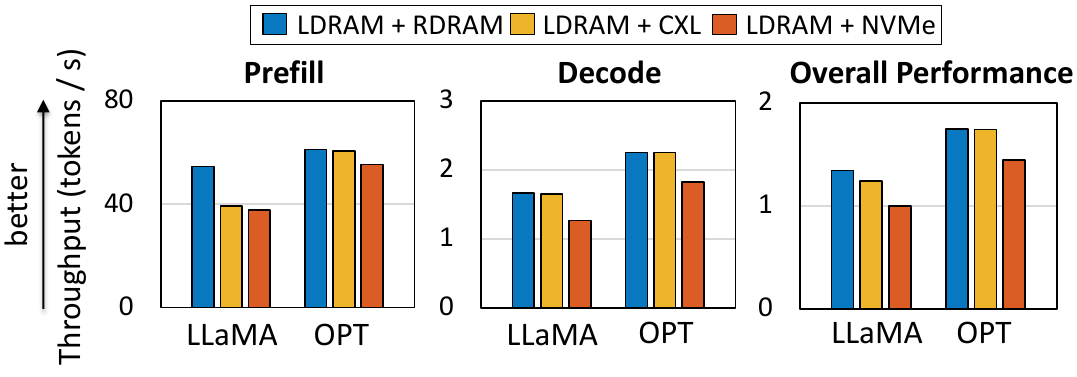}
  \vspace{-25pt} 
  \caption{Comparison LLM inference throughput across memory systems, each with 324 GB capacity.}
\label{fig:diff-media-flexgen}
  \vspace{-15pt}
\end{figure}

\textbf{Evaluation results.}
\underline{LLM inference observation (LIO) 1.}  The performance of the CXL memory is comparable to that of RDRAM, and surpasses NVMe when applied in LLM inference with the tensor offloading.

Figure~\ref{fig:diff-media-flexgen} presents the results. Each evaluation uses two equal-sized memory medias with a total capacity of 324 GB. According to FlexGen's tensor offloading policy, only 8\% of the KV cache resides on the GPU, and the remaining KV cache, weights, and activations reside on the CPU. \textcolor{check}{Figure~\ref{fig:diff-media-flexgen} shows that the inference throughput using LDRAM + CXL is similar to that of LDRAM + RDRAM, with the difference being less than 3\%. Furthermore, LDRAM + CXL shows an improvement of 24\% for LLaMA and 20\% for OPT in overall throughput, compared to LDRAM + NVMe.}

%Figure~\ref{fig:diff-media-flexgen} shows the inference throughput of LLaMA and OPT on host CPU memory systems with various media types. Each system has a memory capacity of \textcolor{red}{324GB}. The tensor offloading policy is determined by FlexGen~\cite{flexgen}, which takes into account the latency and bandwidth characteristics of NVMe. %Details can be found in Table~\ref{tab:flexgen-policy-search}.

%For two-tiered memory systems possessing identical memory capacities, the inference throughput using LDRAM + CXL is similar to that of LDRAM + RDRAM, with the difference being less than 3\%. Furthermore, LDRAM + CXL shows an improvement of 24\% for LLaMA and 20\% for OPT in overall throughput, compared to LDRAM + NVMe.

\underline{LIO 2.} \textcolor{check}{The throughput of prefill and decode stages responds differently to memory latency and bandwidth.} %\textcolor{red}{(should we add this?)}

\textcolor{check}{Figure~\ref{fig:diff-media-flexgen} shows that during the prefill stage, the throughput difference in the prefill stage across the three interleaving policies largely reflects the trend of latency difference in the memory systems. For example,  LDRAM + RDRAM outperforms LDRAM + CXL and LDRAM + NVMe by 20\% and 28\% on average, respectively. This is because during the prefill stage, tensors are loaded from the CPU to the GPU and such data loading is sensitive to the latency.} 

%Specifically, during the prefill stage, tensors are loaded from the host CPU memory systems to the GPU. Consequently, the throughput in the \textcolor{red}{prefill} stage reflects the latency trends of the memory systems, 
%(depicted in Figure~\ref{fig:ai-latency}). 
%as shown in figure~\ref{fig:diff-media-flexgen}, where LDRAM + RDRAM outperforms both LDRAM + CXL and LDRAM + NVMe by \textcolor{check}{20\% and 28\% on average}, respectively.

%%%In the decode stage, attention computation is performed on the CPU, leading to intensive data  access. These operations are particularly sensitive to bandwidth constraints. As a result, the decoding throughput on systems equipped with LDRAM + RDRAM and LDRAM + CXL is similar. The system with LDRAM + CXL is, on average, 27\% more efficient than the system using LDRAM + NVMe. %30.62\% for LLaMA and 23.56\% for OPT 
%%%Moreover, since decoding occurs for each generated token, it predominantly dictates the overall performance. Therefore, the system utilizing LDRAM + CXL outperforms that using NVMe in terms of efficiency.

In contrast, during the decode stage, the throughput is more sensitive to memory bandwidth (see Figure~\ref{fig:diff-media-flexgen}). The decoding throughput using LDRAM + CXL is 27\% better than using LDRAM + NVMe on average. LDRAM + RDRAM and LDRAM + CXL perform similarly. This is because in the decode stage, the attention computation happens on the CPU, leading to intensive data access and sensitive to large bandwidth difference \textcolor{check}{between CXL and NVMe (but not relatively small difference between CXL and RDRAM)}.

\begin{table}[!t]
\centering
%\vspace{5pt}
\caption{LLM inference configuration for evaluation. ``BS'' and ``c'' stands for ``batch size'' and ``KV cache'', respectively. According to offloading policy in FlexGen, all weights and activations are stored in CPU memory. }
\label{tab:flexgen-policy-search}
\vspace{-5pt}
%\resizebox{0.99\textwidth}{!}{
\scriptsize	
\begin{tabular}{|c|c|c|c|c|c|}
\hline
\textbf{LLM} & \textbf{Memory hierarchy} & \textbf{BS}  
 & \textbf{\makecell{\textit{c} on\\GPU }} & \textbf{\makecell{\textit{c} on\\CPU}}  & \textbf{\makecell{Memory\\footprint}} \\ \hline
%LLaMA & LDRAM + NVMe (324 GB) & 30 & 0 & 0 & 100 & 8 & 86 & 6  &  291 GB \\ \hline
%LLaMA & LDRAM + CXL (324 GB) & 30 & 0 & 100 & \textbackslash & 8 & 92 & \textbackslash & 291 GB\\ \hline
%LLaMA & LDRAM + RDRAM (324 GB) & 30 & 0 & 100 & \textbackslash & 8 & 92 & \textbackslash &291 GB \\ \hline
%OPT & LDRAM + NVMe (324 GB) & 28 & 0 & 0 & 100 & 8 & 86 & 6 & 265 GB\\ \hline
%OPT & LDRAM + CXL (324 GB) & 28 & 0 & 100 & \textbackslash & 8 & 92 &  \textbackslash & 265 GB\\ \hline
%OPT &  LDRAM + RDRAM (324 GB) & 28 & 0 & 100 & \textbackslash & 8 & 92 & \textbackslash & 265 GB\\ \hline \hline
LLaMA & LDRAM Only (196 GB) & 14 & 20\% & 80\% &  200 GB\\ \hline
LLaMA &  LDRAM + RDRAM (392 GB) & 40 & 4\% & 96\% &  348 GB \\ \hline
LLaMA & Interleave all (520 GB) & 56 & 4\% & 96\%  & 438 GB \\ \hline
OPT & LDRAM Only (196 GB) & 9 & 27\% & 73\% &  168 GB\\ \hline
OPT &  LDRAM + RDRAM (392 GB) & 40 &  4\% & 96\% & 326 GB\\ \hline
OPT & Interleave all (520 GB) & 64  & 4\% & 96\%  & 448 GB\\ \hline
\end{tabular}
%}
\vspace{-12pt}
\end{table}

\begin{figure}[!tbp]
  \centering
   \includegraphics[width=1.0\columnwidth]{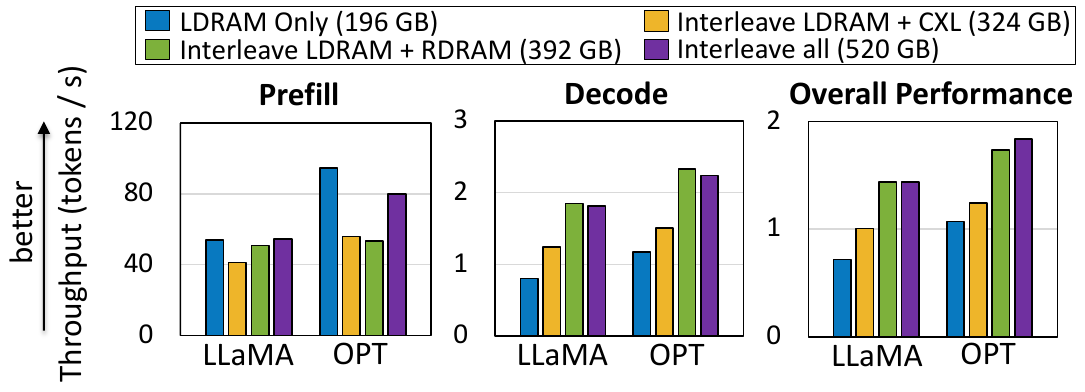}
   \vspace{-20pt}
  \caption{Comparison of LLM inference throughput across various memory systems with different capacities.}
   \vspace{-20pt}
\label{fig:diff-capacity-llama-opt}
  % \vspace{-5pt}
\end{figure}

\underline{LIO 3.} CXL increases the capacity of the memory system, thereby enhancing the throughput of LLM inference \textcolor{check}{due to larger batch size.}
%CXL as capacity expansion can increases LLM inference throughput.

Figure~\ref{fig:diff-capacity-llama-opt} illustrates the inference throughput with various capacities of the CPU memory. \textcolor{check}{Table~\ref{tab:flexgen-policy-search} summarizes the offloading policy search results with those capacities.} The LLM batch size scales up with the increased memory system capacity. Specifically, with LDRAM + CXL, LDRAM + RDRAM, and LDRAM + RDRAM + CXL, the batch size increases by 1.14$\times$, 1.85$\times$, and 3$\times$ for LLaMA, and 2.11$\times$, 3.44$\times$ and 6.11$\times$ for OPT, respectively, compared to the case of utilizing LDRAM only. The overall throughput increases by 28\%, 81\% and 86\% on average respectively, compared to LDRAM only. % 40.11\%, 100.13\%, and 100.41\% for LLaMA, 16.15\%, 61.99\%, and 71.70\% for OPT

\textcolor{check}{We further decompose performance into prefill and decode. During the prefill stage, the inference throughput is dominated by the latency of GPU accessing the CPU memory hierarchy, hence LDRAM only outperforms LDRAM + CXL, LDRAM + RDRAM, and LDRAM + RDRAM + CXL by 50\%, 41\%, and 9\% on average, respectively.} However, during the decode stage, as the batch size increases, the inference throughput improves by \textcolor{check}{42\%, 114\%, and 109\% on average for LDRAM + CXL, LDRAM + RDRAM, and LDRAM + RDRAM + CXL respectively}, compared with LDRAM only. 

%\textcolor{ren}{The throughput of the prefill stage is dominated by the latency of GPU access the host CPU memory system.} Specifically, LDRAM-only outperforms those with LDRAM + CXL, LDRAM + RDRAM, and LDRAM + RDRAM + CXL by \textcolor{check}{50\%, 41\%, and 9\%} on average. %\textcolor{red}{This advantage is due to the high level of parallelism present in the prefill stage.} 
%{30.84\%, 6.29\%, -0.97\% for LLaMA, 68.70\%, 76.68\% and 18.11\% for OPT}

%%%%showing an average rise of \textcolor{red}{42\%, 114\%, and 109\% on average over LDRAM-only} in systems equipped with LDRAM + CXL, LDRAM + RDRAM, and LDRAM + RDRAM + CXL, respectively.  %{55.43\%, 130.83\%, and 126.71\% for LLaMA, 28.47\%, 99.06\% and 91.3\% for OPT}
%%%LLM inference benefits from the extensive memory capacity offered by CXL, as it helps in fully maximizing \textcolor{ren}{the system's computational capabilities.}

\vspace{-5pt}
\begin{tcolorbox}
\footnotesize
\textbf{Takeaway:} (1) The CXL memory as a memory capacity expander enables LLMs to train and infer with larger batch sizes, which in turn improves system throughput. However, when the GPU accesses the CXL memory, the lack of peer-to-peer access support in CXL 1.1 prevents them from leveraging the extra bandwidth and results in longer latency due to the extended data path. (2) Computation offloaded to the CPU can benefit from extra CXL bandwidth.  
\end{tcolorbox}

\section{HPC Workloads}
\label{sec:interleave}

\begin{comment}
\begin{figure}[tb!]
	\centering
	\includegraphics[width=0.88\linewidth]{./figures/all_hpc.cpu1.h.norm.wider.v2.pdf}
	\vspace{-10pt}
 \caption{The overall performance for all benchmarks using various memory components. ``XS'' stands for ``XSBench''.}%\textcolor{red}{((1)change it to the normalized performance. (2) put ``Number of Threads:4 and 32'' at the bottom of the figure. (3) Make the words in the figure larger, as large as the words in paper or slightly bigger. (4) Use ``XS'' or other abbreviations to represent the XSBench to save space?)}}
	\centering
	\label{fig:hpc_overall_perf}
    \vspace{-5pt}
\end{figure}
\end{comment}

We analyze HPC workload performance, focusing on seven workloads from NPB~\cite{NPB} and XSBench benchmark~\cite{tramm2014xsbench}, which require processing a large working memory set using multiple threads, are summarized in Table~\ref{tab:hpc_workloads}. 
They collectively cover the most common and representative HPC applications~\cite{Asanovic+:TR06}. %a diverse range of HPC applications. 
We use various memory allocation policies, including preferred and interleaving. The preferred policy for a specific memory node indicates that the memory is allocated in that memory node first; when that memory node runs out of space, the page allocation goes to another memory node closet to the CPU according to the NUMA distance. The interleaving policy indicates that pages are allocated among memory nodes in a round robin fashion. We present evaluation results on the system A, \textcolor{check}{because we have limited accesses to B and C.} 

\begin{comment}
    We analyze HPC workload performance to demonstrate the use case of CXL memory, focusing on seven workloads from \textcolor{sherry}{NPB~\cite{NPB}} and XSBench benchmark~\cite{tramm2014xsbench}. These workloads, which require processing a large working memory set using multiple threads, are summarized in Table~\ref{tab:hpc_workloads}. They collectively cover the most common and representative HPC applications~\cite{Asanovic+:TR06}. %a diverse range of HPC applications. 
In our study, we use various memory allocation policies, including preferred and interleaving. The preferred policy for a specific memory node indicates that the memory is allocated in that memory node first; when that memory node runs out of space, the page allocation goes to another memory node closet to the CPU according to the NUMA distance. The interleaving policy indicates that pages are allocated among memory nodes in a round robin fashion. We present evaluation results on the system A, because its performance is between the performance of the systems B and C, hence can be representative. \textcolor{dong}{Also, we have limited accesses to B and C.}
\end{comment}

%We quantify the benefits of memory configurations on HPC applications that require large memory capacity. We choose seven popular HPC workloads from NPB benchmark suite~\cite{} and XSBench benchmark suite~\cite{}.  \textcolor{red}{Each application uses several input problems whose memory footprint eventually scale beyond the DRAM capacity. }

\begin{table*}[!t]
\centering
\caption{HPC workloads for evaluation. ``BW'' stands for ``bandwidth''.}
\label{tab:hpc_workloads}
\vspace{-5pt}
\resizebox{0.95\textwidth}{!}{
\begin{tabular}{|c|c|c|c|c|c|}
\hline
\textbf{} & \textbf{Type}         & \textbf{Workload Characterization} &  \multicolumn{1}{l|}{\textbf{Input Problem}} & \textbf{Mem. footprint} &\textbf{BW-hungry Objects} \\ \hline
BT        & Dense linear algebra  &  Unit-strided memory accesses from dense matrices   &  Class E & 166 GB   &  u(39.6G), rsh(39.6G), forcing(39.6G)   \\ \hline
LU        & Sparse linear algebra &  Indexed loads and stores from compressed
matrices  &  Class E  &  134 GB& u(39.6G), rsd(39.6G)     \\ \hline
CG        & Sparse linear algebra &Irregular memory accesses based on indirect indexing  & Class E  &  134 GB & a(48.9G) \\ \hline
MG        & Structured grids      &  Dynamic updates based on subdivided regular grids &  Class E   &  210 GB & v(64.2G), r(73.4G)  \\ \hline
SP        & Structured grids      &   Intense floating-point computations for linear equations  &   Class E  &     174 GB  & u(39.6G), rsh(39.6G), forcing(39.6G) \\ \hline
FT        & Spectral method       &   Bandwidth-consuming matrix transpose  &  Class D & 80 GB &    u0(32.0G), u1(32.0G)  \\ \hline
XSBench   & Monte Carlo           &   Computation based on repeated random trials &  Extra large &  116 GB  & nuclide\_grids \\ \hline
\end{tabular}
}
\vspace{-14pt}
\end{table*}

%%%\textbf{Methodology.}
%%%%\textcolor{red}{depict the interleaving options.}

\subsection{Evaluation Results}
\textcolor{check}{We have the following observations unseen in the existing CXL evaluation~\cite{micro23_cxl,eurosys24_cxl}.} %\textcolor{red}{(I deleted HPC observation 0 to save space. Let me know if it makes sense.)}

\begin{comment}
\underline{HPC observation 0.} Using CXL preferred, some HPC applications show minor performance loss, compared to using LDRAM and RDRAM. 

Figure~\ref{fig:hpc_overall_perf} shows the performance. We observe that BT and LU, when the number of threads is small (4), the performance difference between LDRAM and CXL preferred is very small (less than 3.2\%). These benchmarks are highly compute intensive and not sensitive to long memory latency and low bandwidth with a small number of threads. These results show great potential of using CXL memory to save LDRAM.  For other benchmarks, the performance difference can be up to $1.9\times$ when the number of threads is 4. %%%These benchmarks are either latency sensitive (CG, SP and XSBench) or bandwidth sensitive (MG, FT, BT and LU). 

We also notice that when the number of threads is large (32), the performance difference between LDRAM and CXL preferred is at least $1.9\times$, and the performance difference between RDRAM and CXL preferred is at least $1.1\times$. When the number of threads is large, the memory-level parallelism imposes high pressure on memory components, exacerbating inferior performance of the CXL memory.
\end{comment}

\underline{HPC observation 1.} When the interleaving involves the CXL memory, we can save LDRAM by using RDRAM.

\begin{figure*}[t!]
	\centering
	\includegraphics[width=1.0\linewidth]{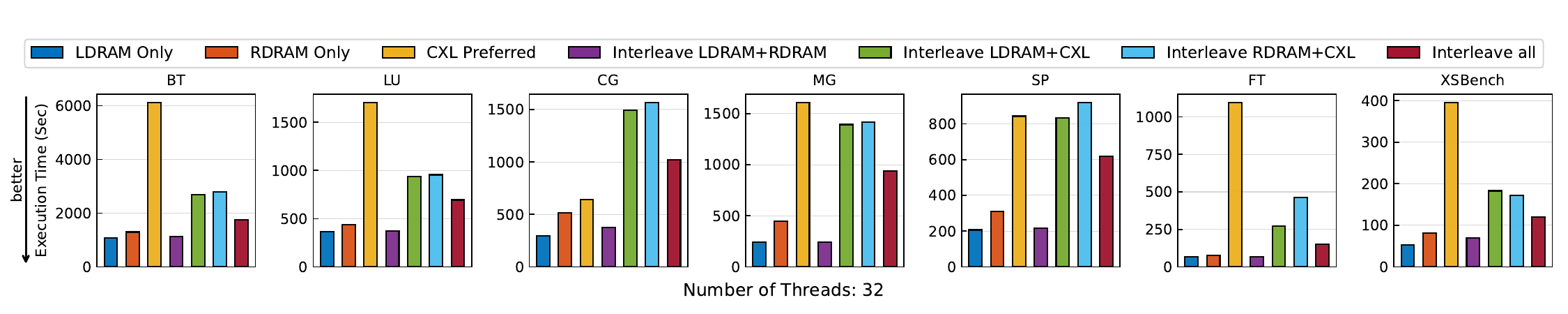}
 \vspace{-25pt}
	\caption{The performance of various interleaving policies for HPC applications.}
 \vspace{-20pt}
	\centering
	\label{fig:hpc_observation3}
\end{figure*}

Figure~\ref{fig:hpc_observation3} shows the results with various interleaving policies, and the benchmarks run on CPU 0. We can see that the performance difference between the interleaving RDRAM+CXL and interleaving LDRAM+CXL is less than $9.2\%$ for all benchmark. 
% In general, \textit{interleaving CXL and RDRAM, we save LDRAM to achieve similar performance.}
In general, \textcolor{revision}{we can save LDRAM while achieving similar performance by interleaving RDRAM+CXL instead of interleaving LDRAM+CXL.}

The above results are because of large performance gap between the CXL memory and DDR (e.g., $2.1\times$ and $1.2\times$ longer in memory access latency, compared to LDRAM and RDRAM respectively.): because of data dependency and limited hardware resources (e.g., Miss Status Handling Registers (MSHR) and the queues in MC), the performance is highly impacted by the slow CXL memory and irrelevant whether the LDRAM or RDRAM is utilized.  

\underline{HPC observation 2.} Bandwidth-sensitive and latency-sensi\\-tive applications respond differently to the bandwidth increase offered by CXL. 

%\textcolor{red}{Add two figures to show MG and CG respectively. The x axis is the number of threads, and the y axis is execution time normalized by the performance with DDR only (one socket). For each thread, there are three bars: local DDR, interleave local DDR and remote DDR (interleave 0 and 1), interleave local DDR and CXL (interleave 0 and 2), interleave all (including CXL, i.e., interleave 0, 1, and 2).}

\begin{figure}[tb!]
	\centering
	\includegraphics[width=1.0\linewidth]{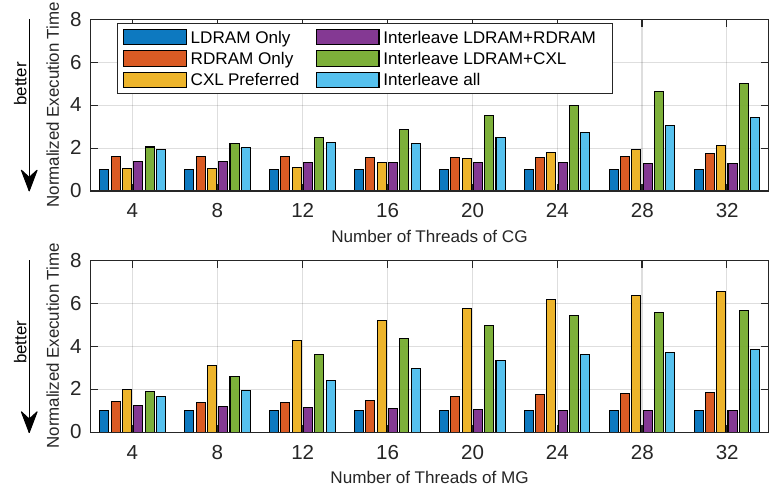}
    \vspace{-20pt}
	\caption{The scalability study for CG and MG. The performance is normalized by that of using LDRAM only.}
	\centering
    \vspace{-15pt}
	\label{fig:hpc_observation2}
\end{figure}

Figure~\ref{fig:hpc_observation2} shows the results for MG (bandwidth-sensitive) and CG (latency-sensitive). For MG, we see that when the number of threads increases from 4 to 32, the performance of ``interleave all'' (i.e., interleaving between LDRAM, RDRAM and CXL, which achieves the highest bandwidth) is consistently better than that of CXL preferred by 10\%-85\%. 
%CXL preferred and interleave all 
%becomes smaller 
%(changing from \textcolor{jie}{%$1.5\times$ to $3.8\times$}),
%10\% to 85\%)}.
Also, for CG, the performance of ``interleave all'' performs worse than that of CXL preferred by up to 1.6$\times$, because using CXL preferred, consecutive memory accesses tend to fall into the same memory node, which is favored by CG for high performance (see more discussions for HPC observation 3).
%%In other words, \textit{the bandwidth benefits offered by CXL transforms into the latency reduction. A similar observation has been shown in high-bandwidth memory~\cite{das2020manage}, but is never seen on the CXL machine.} The latency reduction comes from less contention on MC and hence reducing the waiting time of pending memory transactions on MC. 

%%For \textcolor{red}{CG}, we see the opposite performance: when the number of threads increases from 4 to 32, the performance difference between LDRAM only and interleave all becomes larger (changing from \textcolor{jie}{$1.5\times$ to $3.4\times$}). Such an increase in the latency comes from the longer latency in CXL (compared to LDRAM).

%%%%\textcolor{red}{Add a figure. The x axis is the benchmark names. Each benchmark has three bars: local DDR only, interleave 0 and 1, interleave 0 and 2, interleave 1 and 2, interleave 0, 1, and 2. Use 32 threads.}

%\textbf{Observation 4:} Maximizing the interleave between memory components can be useful for application performance. 

\underline{HPC observation 3.} The CXL memory can show unexpected high performance for the latency-sensitive applications with random memory accesses. 

The \textcolor{check}{CG} results in Figures~\ref{fig:hpc_observation3} and~\ref{fig:hpc_observation2} support the above observation. CG is a latency-sensitive application because of its indirect indexing-based memory accesses. In Figure~\ref{fig:hpc_observation2}, CXL preferred performs better than \textcolor{check}{using RDRAM-only} by $10.9\%$-$57.2\%$ when the number of threads is 4-20. This seems to be counter-intuitive, because the CXL memory has higher latency than RDRAM. %We think that this superior performance may come from the optimization in the CXL device or customized caching policy in the processor 
\textcolor{check}{This superior performance comes from the optimization in the CXL device or customized caching policy in the processor for expensive, CPU-less memory accesses~\cite{micro23_cxl},}  and this optimization is especially effective for CG-style workloads. When the number of threads is larger than 20, the CXL memory accesses become more intensive, and CXL inferior performance becomes more obvious.

%\textcolor{ren}{In Figure~\ref{fig:hpc_observation3}, we also observe that  an very interesting performance behavior of \textcolor{check}{CG} on CXL memory: ``CXL preferred'' outperforms all CXL-related interleaving policies, despite the latter providing higher bandwidth and shorter average latency. For latency-sensitive workloads with random and fine-grained memory access patterns, such as CG, the consistency of access latency becomes more important than the average latency or bandwidth.}

In Figure~\ref{fig:hpc_observation3}, we also observe that CXL preferred outperforms any other CXL-related interleaving policies \textcolor{check}{in CG}, despite other polices provide higher bandwidth and shorter average access-latency. This indicates that for a latency-sensitive workload with a random, scattered memory accesses, gathering accesses in one memory node instead of spreading to multiple memory nodes benefits performance because of reduction of row buffer misses in memory devices.

%{the consistency of access latency becomes more important than the average latency or bandwidth.}

%\textcolor{red}{In Figure~\ref{fig:hpc_observation3}, we notice that \textcolor{check}{CG}'s performance on the CXL memory is interesting: interleave all performs better than the other two interleaving policies. interleave all provides the highest bandwidth. For a latency sensitive workload such as \textcolor{check}{CG}, this bandwidth benefit transforms into the latency benefit, \textcolor{check}{because of less contention on MC and data path.} }

\subsection{Object-Level Interleaving}
\label{sec:object_interleave}
% Instead of generally interleaving pages at the application level (named \textit{application-level interleaving}), 
\textcolor{check}{Instead of generally interleaving pages across the entire application (named \textit{uniform interleaving}),} we propose a method to interleave pages at the data object level. This enables fine-grained control over how pages are interleaved. Since different data objects have different access patterns, using the fine-grained control allows the bandwidth-sensitive object to be accessed with high bandwidth, while the latency-sensitive object is allocated locally for high performance. 

%\textbf{Interleaving method.} We use the two criteria to choose data objects with large bandwidth requirement for interleaving. When those data objects are allocated, they use the interleaving policy.  

\textbf{Interleaving method.} %\textcolor{sherry}{To implement object-level interleaving, we apply \texttt{numa\allowbreak\_alloc\allowbreak\_interleaved\allowbreak\_subset()}~\cite{numactl1} to XSBench and NPB-CPP~\cite{NPB_cpp}, a C++ implementation of NPB benchmark suite, instead of the original Fortran version~\cite{NPB}. Additionally, we use two criteria to select data objects and allocate them using the interleaving policy.}
\textcolor{check}{To implement object-level interleaving, we employ  \texttt{numa\allowbreak\_alloc\allowbreak\_interleaved\allowbreak\_subset()}~\cite{numactl1} in Linux, which allows for the allocation of a data object in an interleaved manner among specific NUMA nodes. We use two criteria to select data objects to use the interleaving policy.}

\begin{itemize}[leftmargin=*,noitemsep,topsep=0pt]
\item \textcolor{check}{The object must have a large memory footprint, which means taking at least 10\% of total memory consumption.}

\item Memory accesses to the object must be intensive; among the data objects that meet the first criterion, we select data objects with the largest number of memory accesses. Multiple data objects may be selected.
%\item Memory accesses to the object must be intensive, which means among those data objects that meet the criteria (1), we choose data objects whose memory accesses are the largest (multiple data objects with the same number of memory accesses can be selected).
\end{itemize}

With high thread-level parallelism, memory accesses to the above data objects are more sensitive to memory bandwidth. Besides the above data objects, the memory allocation for other objects uses the ``preferred'' policy. The last column in Table~\ref{tab:hpc_workloads} summarizes those bandwidth-hungry data objects selected for interleaving.

We evaluate the effectiveness of our object-level interleaving using the following approach. In each test, we run the workload on CPU 0 using both LDRAM (memory node 0) and CXL memory. 
% The LDRAM is either 64GB or 128GB, 
\textcolor{revision}{LDRAM is limited to either 64GB or 128GB by using GRUB mmap,} 
and the memory consumption of all applications exceeding 64GB, which allows us to assess the cases with both sufficient and insufficient LDRAM. The CXL memory is consistently 128GB.  We make the following object-level interleaving observations (abbreviated as OLI observations).

% \begin{figure}[tb!]
% 	\centering
% 	\includegraphics[width=1.0\linewidth]{figures/all_hpc_interleave4-modified.pdf}
%     \vspace{-15pt}
% 	\caption{Performance speedup of various memory allocation strategies compared to ``LDRAM preferred'', using sufficient LDRAM (128GB).}
%  \vspace{-5pt}
% 	\centering
% 	\label{fig:interleaved_obj1}
% \end{figure}

\underline{OLI observation 1.} When the local DRAM is sufficient (with 128 GB), the object-level interleaving performance is similar to that with LDRAM preferred, and consistently outperforms the uniform interleaving in all scenarios. 

\begin{figure}[tb!]
	\centering
	\includegraphics[width=1.0\linewidth]{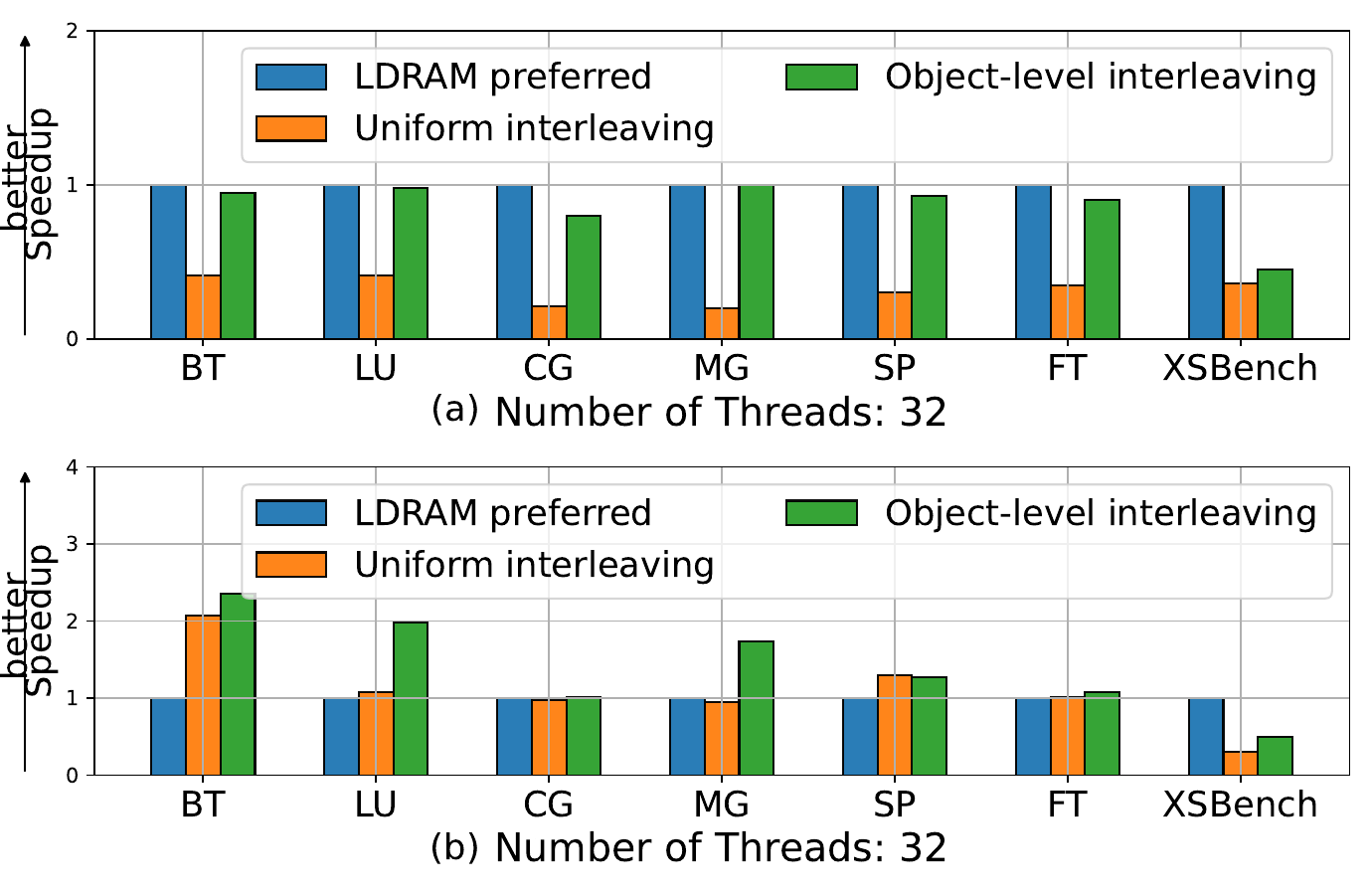}
 \vspace{-25pt}
	\caption{(a) Performance speedup of various memory allocation strategies compared to ``LDRAM preferred'', using sufficient LDRAM (128GB). (b) Performance speedup of various memory allocation strategies, compared to ``LDRAM preferred'', using insufficient LDRAM (64GB).}
 \vspace{-15pt}
	\centering
	\label{fig:interleaved_obj2}
\end{figure}

Figure~\ref{fig:interleaved_obj2}(a) shows that OLI performance is 65\% better than the uniform interleaving on average, demonstrating the effectiveness of OLI to meet the diverse requirements of latency-sensitive and bandwidth-sensitive objects. Additionally, \textit{we observe that by using OLI, we can effectively reduce the fast memory size by an average of %\textcolor{sherry}{\sout{48\% (and up to 80\% for FT)}} \textcolor{sherry}{}
32\% (and up to 40\% for FT), while still achieving similar performance as LDRAM preferred.}  The performance difference between OLI and LDRAM preferred is less than 1\% on average for all HPC workloads, except XSBench, when LDRAM is sufficient. 

% \begin{figure}[tb!]
% 	\centering
% 	\includegraphics[width=1.0\linewidth]{figures/all_hpc_interleave_updated2-modified.pdf}
%   \vspace{-20pt}
% 	\caption{Performance speedup of various memory allocation strategies, compared to ``LDRAM preferred'', using insufficient LDRAM (64GB).}
%  \vspace{-10pt}
% 	\centering
% 	\label{fig:interleaved_obj2}
% \end{figure}

\underline{OLI observation 2.} When the local DRAM is insufficient (64 GB), the OLI outperforms any other cases. 

Figure~\ref{fig:interleaved_obj2}(b) shows that on average, OLI performs 1.42$\times$ better than LDRAM preferred (and up to 2.35$\times$ for BT). It also outperforms the uniform interleaving by 1.32$\times$ on average (and up to 1.84$\times$ for LU). See the following reasons. (1) The performance of LDRAM preferred is highly related to when the data objects are allocated.  If LDRAM is full, latency-sensitive objects might end up in the slower CXL memory. %Those latency-sensitive data objects can be allocated to slow CXL memory when there is no space in LDRAM. 
(2) The bandwidth-sensitive data objects do not have opportunities to be allocated on both LDRAM and CXL for high bandwidth. 

In XSBench, LDRAM preferred performs better than both the uniform and object-level interleaving. This is because most of the memory accesses in XSBench are concentrated in a small, latency-sensitive memory set.

\vspace{-5pt}
%\noindent\fbox{\begin{minipage}{23.5em}
\begin{tcolorbox}
\footnotesize
\textbf{Takeaway}: The CXL memory offers additional bandwidth, but HPC workloads may not benefit from it directly. The uniform interleaving can undermine performance, while the data object-level interleaving delivers performance comparable to or better than the default LDRAM-centric page allocation  while saving fast memory size.
%initialization strategy.
\end{tcolorbox}
%\end{minipage}}

\section{Memory Tiering based on Page Migration}
%This section investigates the performance impact of page migration in the operating system on CXL-based multi-tiered memory. We conduct all evaluations on System A in table~\ref{tab:system_config}.
Memory tiering is an application-transparent solution to integrate the CXL memory into the existing memory systems. Treating the CXL memory as a memory tier, existing memory tiering solutions~\cite{autonuma, tiering0.8, eurosys24:mtm, 10.1145/3582016.3582063, Raybuck2021HeMemST, 10.1145/3600006.3613167} rely on memory profiling to count memory accesses at the page level. %\textcolor{sherry}{(consider to describe as "rely on access tracking at page level")}. 
Then, those solutions move frequently accessed (hot) pages to the fast memory tier, and demote less frequently access (cold) pages to the slow memory tier. In contrast to the interleaving studied in Sections \ref{sec:llm}-~\ref{sec:interleave}, which are static page placement solutions, the memory tiering solutions are dynamic.

\begin{comment}    
In this section, we study the performance differences among various page management policies when applied on System A (Table~\ref{tab:system_config}) to figure out the following questions:
\begin{itemize}[leftmargin=*,noitemsep,topsep=0pt]
    \item How effective are different page migration systems?
    \item How effective are different page placement policies?
    \item Which page management policy is optimal for a CXL-enabled memory system?
\end{itemize}
\end{comment}

%\textbf{Page migration on tiered memory systems.}  CXL enables a tiered memory system with different latency and bandwidth characteristics compared to traditional DRAM-based NUMA systems. In previous sections, we studied various interleaving strategies, which serve as static page placement solutions, to efficiently and transparently leverage the additional bandwidth provided by CXL. In this section, we further explore the impact of dynamic page migration solutions on CXL-enabled tiered memory systems and their interaction with interleaving strategies.

We use the system A. 
We limit the capacity of the fast memory (i.e., LDRAM)\textcolor{revision}{, while the capacity of the slow memory (i.e., CXL) remains the same.}
%to  \textcolor{sherry}{less than half of the application's memory footprint},
% while the slow memory (i.e., CXL) has unlimited capacity.

We evaluate three state-of-the-art memory tiering solutions as follows. We also evaluate \texttt{No Balance} (between NUMA nodes) representing the static page placement without migration. \textit{Our study is featured with analyzing the interplay between page migration and interleaving.} 
%with \texttt{No Balance}  representing the static page placement strategy without migration. %%\textcolor{red}{todo: define promotion/demotion?}

\begin{itemize}[leftmargin=*,noitemsep,topsep=0pt]
% \item \texttt{AutoNUMA}~\cite{autonuma}  tracks memory accesses by unmapping pages to generate hint faults, and periodically scans the application address space to find hint faults. %AutoNUMA identifies hot pages using a predefined, \textit{static} hint fault latency threshold, and promotes pages based on page access recency.
% \textcolor{check}{AutoNUMA promotes pages using a NUMA-distance-based, \textit{static} memory-access threshold to avoid unnecessary page migration.} AutoNUMA is enabled by setting \textit{$numa\_balancing$} to 1 in /proc/sys/kernel (the default Linux setting).

\item \texttt{AutoNUMA}~\cite{autonuma} \textcolor{check}{is the default NUMA-balancing policy in Linux. %It balances overall system load across NUMA nodes. To count memory accesses, \texttt{AutoNUMA} manipulates a specific access bit in PTE such that whenever the page is accessed, an NUMA hint fault is triggered. By capturing the hint faults, \texttt{AutoNUMA} decides where memory accesses come from and relocates memory pages to minimize the distance between the pages and compute. 
\texttt{AutoNUMA} counts memory accesses by manipulating an access bit in PTE. When a page is accessed, a NUMA hint fault is triggered. By tracking these faults, \texttt{AutoNUMA} determines the origin of memory accesses and relocates pages to minimize their distance from the computing processes. 
\texttt{AutoNUMA} is enabled by setting \texttt{numa\_balancing} to 1 in \texttt{/proc/sys/kernel}}.
%\textcolor{sherry}{is a default NUMA-balancing policy in  Linux. It balances the overall system load across NUMA nodes by periodically scanning the application address space. Memory accesses on scanned pages trigger hint faults, prompting \mbox{AutoNUMA} to relocate tasks and their associated memory pages to minimize the distance between tasks and the memory they access.  This process reduces memory access latency and optimizes task efficiency.  \mbox{AutoNUMA} is enabled by setting \textit{$numa\_balancing$} to 1 in /proc/sys/kernel (the default Linux setting).}

% \item \texttt{Tiering-0.8}~\cite{tiering0.8} also uses hint faults to identify hot pages. Unlike AutoNUMA, Tiering-0.8 \textit{dynamically} adjusts the promotion criteria to throttle migration traffic. It considers the recency of page accesses to decide page promotion and demotion. Tiering-0.8 is enabled by setting \textit{$numa\_balancing$} to 2, and it uses the Linux default promotion threshold for initialization. %\textcolor{ren}{todo: distinguish promotion and demotion hotness identification difference. }

\item \texttt{Tiering-0.8}~\cite{tiering0.8} \textcolor{check}{is a recent Linux patch which uses the similar hint fault-based profiling as \texttt{AutoNUMA}. Unlike \mbox{AutoNUMA}, it considers the recency of page accesses based on the re-fault interval to identify hot pages. Additionally, Tiering-0.8 dynamically adjusts the page promotion criteria to throttle migration traffic and save memory bandwidth.  Tiering-0.8 is enabled by setting \texttt{numa\_balancing} to 2.}

%\texttt{Tiering-0.8}~\cite{tiering0.8} \textcolor{sherry}{is a Linux patch from Intel and also migrates pages based on hint faults. Unlike \mbox{AutoNUMA}, it considers the recency of page accesses based on re-fault interval to identify hot pages. Additionally, Tiering-0.8 dynamically adjusts the promotion criteria to throttle migration traffic.  Tiering-0.8 is enabled by setting \textit{$numa\_balancing$} to 2 to support memory tiering system.}

\item \texttt{TPP}~\cite{10.1145/3582016.3582063} uses  hint faults to determine page migration. Upon encountering hint faults, TPP decides to promote a page based on its presence on the LRU list in Linux.

%\item \texttt{TPP}~\cite{10.1145/3582016.3582063} \textcolor{sherry}{also enhances support for page migration within memory tiering system. It utilizes both hint faults and the active LRU list maintained by the OS to determine page migration. Specifically, upon encountering hint faults, TPP decides on the promotion of a page based on its presence on the active LRU list.}
\end{itemize}

\begin{figure*}[t!]
	\centering
	\includegraphics[width=0.9\linewidth]{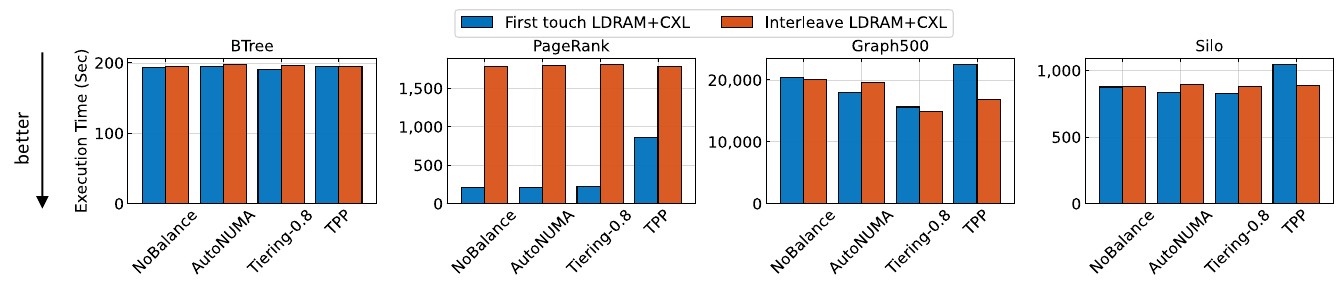}
    \vspace{-15pt}
	\caption{Execution time with various page migration and static page placement solutions for memory-intensive applications.}
    \vspace{-5pt}
	\centering
	\label{fig:pg_time}
\end{figure*}
\begin{figure*}[t!]
	\centering
	\includegraphics[width=1\linewidth]{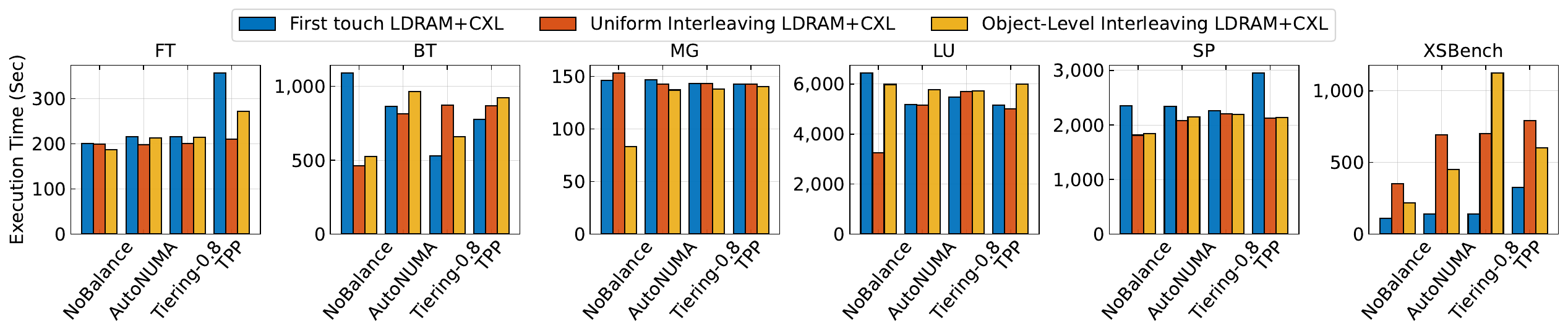}
    \vspace{-22pt}
	\caption{Execution time with various page management solutions for HPC applications.}
	\centering
	\label{fig:pg_hpc_time}
    \vspace{-10pt}
\end{figure*}

\begin{table*}[h]
\centering
\caption{\textcolor{revision}{Comparison of Performance Studies on Genuine CXL}}
\vspace{-5pt}
\label{tab:related_work}
\setlength{\tabcolsep}{7pt}
\begin{tabular}{ccccccc}
\hline
\multirow{2}{*}{\textbf{Paper}} & \multirow{2}{*}{\textbf{HPC Study}} & \multicolumn{2}{c}{\textbf{LLM Study}} & \multicolumn{2}{c}{\textbf{Memory Tiering}} & \multirow{2}{*}{\textbf{Performance Optimization}} \\ \cline{3-6}
                                &                                      & \textbf{Training} & \textbf{Inference} & \textbf{\makecell{Solutions \\ Evaluated}} & \textbf{\makecell{Identify \\Limitations}} &                                                                \\ \hline
MICRO23~\cite{micro23_cxl}                         & No                                   & N/A               & N/A                & TPP                          & N/A                             & BW-aware page allocation                                       \\
EuroSys24~\cite{eurosys24_cxl}                       & No                                   & N/A               & CPU-based          & AutoNUMA                     & N/A                             & N/A                                                            \\
\textcolor{revision}{SupMario~\cite{microsoft24cxl}}                       & \textcolor{revision}{Yes}                                   & \textcolor{revision}{N/A}               & \textcolor{revision}{CPU-based}          & \textcolor{revision}{TPP}                     & \textcolor{revision}{Yes}                             &  \makecell{\textcolor{revision}{Predictive interleaving \&}\\ \textcolor{revision}{regulated migration}}                                                            \\
\textbf{This work}                      & \textbf{Yes}                                  & \textbf{GPU-based}         & \textbf{GPU-based}          & \textbf{TPP}, \textbf{AutoNUMA}, \textbf{Tiering-0.8}   & \textbf{Yes}                             & \textbf{Object-level interleaving}                                      \\ \hline
\end{tabular}
\vspace{-18pt}
\end{table*}

%\subsection{Evaluation Setup}
\subsection{Page Migration with \textcolor{check}{Uniform interleaving}}
\textbf{Evaluation setup}. To study the impact of page migration, we evaluate four memory-intensive applications, including BTree~\cite{btree}, an in-memory index lookup; PageRank~\cite{beamer2015gap} and Graph500~\cite{graph500}, both graph processing applications; and Silo~\cite{tu2013speedy}, an in-memory database engine. We run them with 64 threads.
%to intensify memory traffic and maximize bandwidth consumption. 
We configure the memory consumption of each application to be around 130GB. This configuration ensures a fair comparison between static page placement solutions (i.e., the NUMA first touch and interleaving): LDRAM (i.e., 50GB) is set to less than half of each application's memory consumption, hence both solutions can fully utilize LDRAM.

\textbf{Metrics.}  We collect the execution time and page migration statistics, including the number of hint faults and migrated pages. Such  statistics is collected by periodically reading the Linux counters from  \texttt{/proc/vmstat}. Those counters capture page accesses from the entire system, but primarily influenced by page migrations because of the application.
%\textcolor{dong}{and captures page accesses from both kernel and user spaces (but dominated by the user space).}

%We report the execution time for different applications when running under different page management polices. Furthermore, we also report the number of migrated pages and numa hint faults during the runtime of each applications. This information is calculated by periodically scan the \texttt{/proc/vmstat} file under the Linux system.

%\subsection{Page migration with application-level interleave}

\textbf{Evaluation results.} 
\underline{Page migration observation (PMO) 1.} Different applications perform differently with different page migration and static page placement solutions. No single solution can get the best performance for all applications. \textcolor{check}{The effectiveness of a solution to an application depends on the distribution of hot pages in the working set (e.g., scattered or concentrated), and variance and size of the hot page set.}

% of of applications
%Different applications have varying preferences for combinations of page placement and page migration strategies to achieve optimal performance. 

%Figure~\ref{fig:pg_time} shows the execution time. We observe that BTree is not sensitive to any specific page placement or migration strategies, maintaining relatively stable performance with variations less than \textcolor{red}{1\%}. For PageRank, using the first touch without page migration achieves the best performance, which is \textcolor{red}{7 $\times$} better than interleaving with various page migration strategies. For Graph500, interleaving with Tiering-0.8 outperforms other cases by up to \textcolor{red}{25\%}. For Silo, the optimal performance is achieved using first touch with Tiering-0.8, outperforming other cases by up to \textcolor{red}{15\%}. Such differences arise from various factors, such as memory access patterns, application implementation, and data structures used. The diverse nature of applications and their memory requirements makes it challenging to find a single page management solution (i.e., placement + migration) that works optimally for all cases. 

Figure~\ref{fig:pg_time} shows execution time. We observe that BTree is not sensitive to any solution (the performance variance is less than 3\%),  because of its irregular memory access patterns. PageRank achieves the best performance with the first touch without page migration, which is 88\% better than any page migration solution plus interleaving, because PageRank has a small and stable set of hot pages. Graph500 achieves the best performance with Tiering-0.8 plus  interleaving, outperforming other solutions by up to 33\%, because Graph500's hot pages are scattered across memory tiers, and interleaving is helpful for improving data locality.
%\textcolor{dong}{because Graph500's hot page set changes from time to time, making page migration effective, meanwhile page interleaving avoids some page migrations by locating pages close to the compute.} % in next observation we will say page migration not working. So we should not highlight migration here.
Silo achieves the best performance with Tiering-0.8 plus the first touch, outperforming other solutions by up to 20\%. %\textcolor{sherry}{\sout{Different from Graph500 where the hot data is distributed to many pages, Silo's hot data is gathered into fewer pages, making the first touch more effective than the interleaving. }} 
%\textcolor{sherry}{Unlike Graph500, Silo employs a B-tree-like structure to implement a cache-friendly design and presents a sparse and regular access pattern, making the first touch more friendly to cache efficiency than the interleaving. }
Silo implements a B-tree-like data structure gathering hot data into fewer pages, making the first touch more effective than interleaving.
%%%%\textcolor{red}{(Pending: summarize the key points in the observation.)}

%\textit{We also find that Tiering-0.8 is very effective for memory tiering, compared with TPP and AutoNUMA.}} 
%Such differences arise from various factors, such as memory access patterns, application implementation, and data structures used. The diverse nature of applications and their memory requirements makes it challenging to find a single page management solution (i.e., placement + migration) that works optimally for all cases. 

\underline{PMO 2.} When using first touch, Tiering-0.8 outperforms TPP and AutoNUMA, because of its smaller profiling overhead and dynamic adjustment of the page promotion threshold.
%provides the largest performance benefit among all  solutions.  

%Figure~\ref{fig:pg_time} illustrates that, using first touch and averaging across various applications, Tiering-0.8 outperforms \texttt{NO Balance}, AutoNUMA, and TPP by average margins of \textcolor{red}{24\%, 13\%, and 75\%, respectively.} Table~\ref{tab:pagemigration} reveals the reasons by showing the number of hint faults and pages migrated with different configurations. Compared to TPP, Tiering-0.8 has \textcolor{red}{xxx $\times$ fewer hint faults on average}. Memory profiling in TPP incurs a large overhead, and the reduced number of hint faults in Tiering-0.8 indicates less profiling overhead, contributing to improved overall performance. Compared to AutoNUMA, Tiering-0.8 has a comparable number of hint faults but \textcolor{red}{xxx $\times$ fewer} pages migrated on average. Despite the lower number of pages migrated, Tiering-0.8 still outperforms AutoNUMA, indicating that the dynamic promotion threshold effectively identifies the most critical pages for migration.

Figure~\ref{fig:pg_time} illustrates that using the first touch, Tiering-0.8 outperforms \texttt{NO Balance}, AutoNUMA, and TPP by 7\%, 3\%, and 31\%. %Table~\ref{tab:pagemigration} reveals the reasons. 
%\textcolor{dong}{Our analysis shows that compared to TPP},
\textcolor{check}{We quantify page hints faults caused by different memory tiering strategies and observe that} Tiering-0.8 has 59 $\times$ fewer hint faults. Memory profiling in TPP incurs a large overhead, and the reduced number of hint faults in tiering-0.8 contributes to performance improvement. %indicates less profiling overhead, contributing to performance improvement. 
Compared to AutoNUMA, Tiering-0.8 has a comparable number of hint faults %and \textcolor{ren}{achieves 50 $\times$ more} pages migrated on average, 
but exhibits large variance in the number of migrated pages. For example, in PageRank, Tiering-0.8 results in 7.6 million more migrations. %%than AutoNUAM. 
This variance is due to adaptiveness of page promotion threshold in Tiering-0.8.

%\textcolor{sherry}{Despite the smaller number of pages migrated, Tiering-0.8 still outperforms AutoNUMA, indicating that the dynamic promotion threshold in Tiering-0.8 effectively identifies the most critical pages for migration. (sherry: rewrite it to :This indicates that the aggressive page demotion in Tiering-0.8 effectively frees up space in the fast memory for page promotion. Additionally, the dynamic promotion threshold in Tiering-0.8 enables it to outperform AutoNUMA.)}

%\underline{PMO 3.} \textcolor{red}{Application level interleave does not function orthogonally with page migration strategies based on hint faults.}

\underline{PMO 3.} \textcolor{check}{The page migration based on hint faults is not integrated well with the uniform interleaving.}

%Table~\ref{tab:pagemigration} reveals that when page migration is combined with interleave, the number of hint faults generated is \textcolor{red}{on average 1000 $\times$ less} than that of page migration combined with first touch. Our further investigation into the Linux kernel reveals that when explicit page placement policies are specified, such as using application-level interleave, application pages are placed in a protected\textcolor{sherry}{(replace to unmigratable)} region. This protection prevents these pages from triggering hint faults. \textcolor{red}{We noticed there even with much less number of hint faults, due to page migration in other applications in systems.}

%Table~\ref{tab:pagemigration} 
%\textcolor{dong}{Our analysis} reveals that 
\textcolor{jie}{We observe that }using page migration plus the \textcolor{check}{application level} interleaving, the number of hint faults is 72,721 $\times$ less than that of page migration plus the first touch on average.  \textcolor{check}{Our investigation reveals that when the application-level interleaving is employed, the application's pages are placed in unmigratable regions, preventing the pages to trigger hint faults. Hence, page migration cannot be effective.} %%%%\textcolor{red}{We noticed there even with much less number of hint faults, due to page migration in other applications in systems.}}

\subsection{Page Migration with Object-Level Interleaving}
\label{sec:obj_interleaving}
%To study page migration plus object-level interleaving, we evaluate HPC workloads listed in Table~\ref{tab:hpc_workloads}. In our experimental setup, the LDRAM is configured with a capacity of 50GB, while the memory consumption on CXL memory is not limited. All HPC workloads are executed using 32 threads on a single CPU socket.
%\textbf{Evaluation setup.} We evaluate HPC workloads listed in Table~\ref{tab:hpc_workloads}.
\textbf{Evaluation setup.} %\textcolor{sherry}{We evaluate the C++-implemented HPC workloads listed in Table~\ref{tab:hpc_workloads}, consistent with \S\ref{sec:object_interleave}}. 
To enable fair comparison between the \textcolor{check}{uniform} and object-level interleaving, the capacity of LDRAM is set to 40GB for FT, 100GB for MG, and 50GB for others, \textcolor{check}{such that all static page placement solutions can fully utilize LDRAM for each application.} The CXL memory does not have a capacity constraint, \textcolor{check}{because it is the slowest memory tier}. We use 32 threads on the socket 1. 
%\textcolor{sherry}{for a fair comparison between application-level interleaving and object-level interleaving, the capacity of LDRAM is set to 40GB for FT, 100GB for MG, and 50GB for others}, 

%%%All HPC workloads are executed using 32 threads on a single CPU socket.

%The second part of evaluation aims to study the benefit of page migration with object-level interleave. We use 32 threads to run HPC applications in (Table~\ref{tab:hpc_workloads}) and maintain the size of input problem and the interleaved objects same as (Table~\ref{tab:hpc_workloads}).

\textbf{Evaluation results.} \underline{PMO 4.} Using the page migration plus the object-level interleaving results in minor performance improvement on HPC workloads, compared to the object-level interleaving without page migration.  %%%or even a negative impact on HPC workloads. 
%\underline{PMO 3.} Object Interleave policy is less efficient with Page Migration Systems.

%%%%Figure~\ref{fig:pg_hpc_time} shows that without the page migration, the object-level interleaving outperforms both first touch and application-level interleaving by up to \textcolor{red}{xxx\%}. However, page migration negatively impacts workloads with object-level interleaving: when using AutoNUMA, Tiering-0.8, and TPP, performance degrades by \textcolor{red}{xx\%, xx\%, and xx\% on average} respectively, compared to no balance. This occurs because object-level interleaving without migration can enhance application performance by better utilizing the additional bandwidth available in CXL for bandwidth-hungry objects. However, dynamic page migration undermines this benefit by disrupting the initial placement of bandwidth-hungry objects.

Figure~\ref{fig:pg_hpc_time} shows that \textit{without page migration}, the object-level interleaving outperforms the first-touch and \textcolor{check}{uniform interleaving} by up to 45\%. However, page migration negatively impacts effectiveness of the object-level interleaving: when using AutoNUMA, Tiering-0.8, and TPP, the performance degrades by 46\%, 88\%, and 63\% on average, compared to using the object-level interleaving without page migration. This occurs because the object-level interleaving utilizes the additional CXL bandwidth for bandwidth-hungry objects. However, the page migration undermines this benefit. %through page migration. 

%%From the performance comparison of HPC applications in Figure~\ref{fig:pg_hpc_time}, the Object Interleave policy is shown to consistently outperform the Firsttouch or Interleave policies under the NoBalance mode. However, when applications utilize the Object Interleave policy under page migration systems, i.e., AutoNUMA, Tiering-0.8, and TPP, their performance degrades compared to the NoBalance mode. Moreover, when page migration is enabled, the application performance under the Object Interleave policy becomes worse than that under the Firsttouch and Interleave policies. This suggests that while the Object Interleave policy can enhance application performance by providing additional bandwidth to bandwidth-sensitive objects, the advantages it offers are undermined by the effects of page movement.

%\underline{PMO 5.} The page migration can enhance the performance of applications with hot regions while potentially harming others.
%\underline{PMO 5.} Page migration can benefit the performance of applications with hot regions while adversely affecting others. 

\underline{PMO 5.} \textcolor{check}{The page migration can improve performance for some applications but lose performance for others.} 

%As illustrated in Figure~\ref{fig:pg_hpc_time}, different HPC workloads exhibit varying preferences for page migration. For example, page migration degrades performance for FT, SP and XSBench, and yields almost no performance difference in MG, regardless of the page placement strategy employed, because these workloads only have uniformly accessed memory region or highly skewed hot memory region which make hotness detection very challenging. In contract, page migrations helps improve the performance of BT and LU by up to \textcolor{red}{xx and xx}, respectively, since BT and LU have hot page regions which are easy to identify.  

Figure~\ref{fig:pg_hpc_time} shows that different HPC workloads exhibit different preferences for page migration. For example, the page migration degrades performance for FT, SP and XSBench, and yields almost no performance difference in MG, regardless of using the  interleaving or first touch, because these workloads have uniformly accessed working set or highly skewed and \textcolor{check}{scattered} hot memory region, which make hotness detection challenging. In contrast, the page migration improves BT and LU performance by up to 51\% and 20\%, respectively, since hot pages in BT and LU have good locality to be detected.

%Figure~\ref{fig:pg_hpc_time} shows that FT's execution times for all page placement policies (Firsttouch, Interleave, Object Interleave) under NoBalance are consistent, ranging from 187 to 201 seconds. Introducing page migration systems degrades FT's performance across all policies compared to NoBalance. This performance consistency under NoBalance suggests FT lacks significant hot regions, making it insensitive to page placement, and that page migration can harm performance for such applications.
%In contrast, applications like BT, MG, SP, and LU show significant performance variance with different page placement policies under NoBalance, indicating potential hot memory access regions. These applications, particularly with the Firsttouch policy, can benefit from page migrations. Specifically, the performance of BT, MG, LU, and SP with Firsttouch is improved by up to 51\%, 2\%, 20\%, and 4\%, respectively. XSBench is an exception, where performance under Firsttouch is significantly better than under Interleave and Object Interleave across all page migration systems. This is attributed to XSBench having a highly skewed hot memory region in the early allocated memory area. Therefore, applications with hot regions can leverage page migration systems to improve performance, whereas those without hot regions may experience performance downgrades.

\begin{tcolorbox}
\footnotesize
\textbf{Takeaway:} (1) There is significant potential to improve the performance of page migration in the memory tiering solutions. (2) Dynamic page migration and static page interleaving are not well-integrated. (3) Dynamic page migration may degrade performance; a better static page placement strategy \textit{without page migration}, such as object-level interleaving, can lead to better performance. 
\end{tcolorbox}

\section{Related Work}
\label{sec:related_work}
% \textbf{CXL} has garnered significant attentions from academia and industry~\cite{sharma2023introduction, li2023pond, 10.1145/3582016.3582063, 288782, 288786, shan2022towards, gouk2023memory, zhang2023partial, wahlgren2022evaluating, Sim2023ComputationalCS, cho2023case, li2023understanding, berger2023design,lerner2024cxl,song2023lightweight,alonso2023rethinking }.
% Recent studies analyzed the performance of actual CXL hardware~\cite{micro23_cxl, eurosys24_cxl}. Our work differs in multiple perspectives, summarized in Table~\ref{tab:related_work}.

\textcolor{revision}{\textbf{CXL.} In 2019, Intel introduced Compute Express Link (CXL~\cite{van2019hoti, sharma2022compute}),  an open industry-standard interconnect between processors and devices such as accelerators, memory buffers, and smart network interfaces. Since its introduction, the CXL has garnered significant attention and investment from both researchers and industry practitioners~\cite{sharma2023introduction, li2023pond, 10.1145/3582016.3582063, micro23_cxl, 288782, 288786, shan2022towards, gouk2023memory, zhang2023partial, wahlgren2022evaluating, Sim2023ComputationalCS, cho2023case, li2023understanding, lerner2024cxl, song2023lightweight, alonso2023rethinking, levis2023case, sc24_cxl}. For example, Google has explored the potential of CXL memory in memory tiering and swapping in its cloud computing infrastructure~\cite{levis2023case}, while Microsoft has developed CXL-based memory pools for public cloud platforms~\cite{berger2023design, li2023pond}, and Meta has designed the CXL-based tiered memory system in hyperscale datacenters~\cite{10.1145/3582016.3582063}.}

% \textcolor{revision}{\textbf{CXL} has garnered significant attentions from academia and industry~\cite{sharma2023introduction, li2023pond, 10.1145/3582016.3582063, 288782, 288786, shan2022towards, gouk2023memory, zhang2023partial, wahlgren2022evaluating, Sim2023ComputationalCS, cho2023case, li2023understanding, berger2023design,lerner2024cxl,song2023lightweight,alonso2023rethinking }.} %%%%%%%For example, Google has explored the potential of the CXL memory in memory tiering and swapping in its cloud computing infrastructure~\cite{levis2023case}, while Microsoft has developed CXL-based memory pools for public cloud platforms ~\cite{berger2023design, li2023pond}. 
 %For example, Yang et al.~\cite{288786} overcome the memory wall with emulated CXL-enabled SSDs. 
%There is also a surge in research focusing on specific applications of CXL memory~\cite{berger2023design, cho2023case, lerner2024cxl, song2023lightweight, li2023understanding, alonso2023rethinking, 288782}. 
% \textcolor{revision}{Most of the explorations of using the CXL memory leverage NUMA servers to emulate CXL memory performance~\cite{yang2023cxlmemsim, 288786, 288782, gouk2023memory}.}
\textcolor{revision}{\textbf{Emulation-based CXL study.}}
\textcolor{revision}{Most of the explorations of using the CXL memory from both industry and academics leverage NUMA servers to emulate CXL memory performance~\cite{yang2023cxlmemsim, 288786, 288782, gouk2023memory,wu2024performance}. For example, Yang et al.~\cite{288786} overcome the memory wall with emulated CXL-enabled SSDs, Jang et al.~\cite{288782} apply emulated CXL memory to the billion-scale approximate nearest neighbor search using a software-hardware co-design solution.}

\textcolor{revision}{\textbf{Performance study on genuine CXL.}} 
\textcolor{revision}{Recent studies analyzed the performance of actual CXL hardware~\cite{micro23_cxl, eurosys24_cxl, microsoft24cxl}.}
\textcolor{revision}{Sun et al.~\cite{micro23_cxl} represent the first comprehensive effort to analyze CXL memory performance using genuine CXL-ready systems and devices. Nonetheless, a wide range of applications from some fields (such as HPC, AI inference and training) are still not extensively studied. Tang et al.~\cite{eurosys24_cxl} investigate the CXL performance in various datacenter scenarios and propose an Abstract Cost Model to estimate the cost-benefit of using CXL memory.  However, their work does not address the limitations of existing memory tiering solutions.  Liu et al.~\cite{microsoft24cxl} analyze the root causes of CXL latency's impact on system performance. They further propose a performance model designed to predict CXL-induced slowdowns, guide the uniform interleaving ratio, and regulate page migration throttling in memory tiering. Nonetheless, their performance evaluation is restricted to CPU-centric scenarios.}
\textcolor{revision}{Our work differs in multiple perspectives, as shown in Table~\ref{tab:related_work}: (1) workload diversity, covering both LLMs and HPC; (2) practicality of the evaluation platform, which utilizes GPUs instead of CPUs for LLM; (3) deep insights into memory tiering; and (4) performance optimization techniques, such as object-level interleaving (Sec.~\ref{sec:obj_interleaving}) and thread assignments based on bandwidth scaling (Sec.~\ref{sec:basic_perf}).}

%Recent work studies the performance of real CXL hardware~\cite{micro23_cxl, eurosys24_cxl}. Our work is different from them in terms of (1) performance optimization techniques (e.g., object-level interleaving~\ref{sec:obj_interleaving} and thread assignments based on bandwidth scaling (Section~\ref{sec:basic_perf}) in this paper); (2) workload practicality (e.g., using GPU instead of CPU for LLM training \textit{and} inference); (3) insights on memory tiering solutions. Table~\ref{tab:related_work} summarizes the difference. 

%Besides the above efforts, Sun et al.~\cite{micro23_cxl} demystify the CXL memory performance using genuine CXL-ready systems and devices, which is the first paper that explores  the real CXL-ready systems. However, the vast of applications from other fields (such as HPC and AI) are not extensively studied, and memory tiering and its interplay with page interleaving are not studied. 

%In this paper, we present a comprehensive analysis for CXL-ready systems use three hard CXL IPs from different vendors. We explored how the representative HPC applications~\cite{tramm2014xsbench, NPB_cpp, beamer2015gap, btree, tu2013speedy}, large language model (LLM) inference (LLaMA~\cite{touvron2023llama}, and OPT~\cite{zhang2022opt}), and LLM~\cite{devlin2018bert, radford2019language} training via tensor offload strategy~\cite{Ren2021ZeROOffloadDB} can be benefit from using the CXL memory. 

\textcolor{revision}{\textbf{Tiered memory systems.} }%%%%Different memory components %%%~\cite{wang2020characterizing, xiang2022characterizing, bergman2022reconsidering, dragojevic2014farm, gu2017efficient, abulila2019flatflash, van2019hoti, sharma2022compute} 
%%%%can show different memory latency, bandwidth, capacity, and monetary cost. 
\textcolor{revision}{Different memory components~\cite{wang2020characterizing, xiang2022characterizing, bergman2022reconsidering, dragojevic2014farm, gu2017efficient, abulila2019flatflash, van2019hoti, sharma2022compute,gpu_dp_micro14} can show different memory latency, bandwidth, capacity, and monetary cost. Using multiple memory components, tiered memory systems can save cost and improve memory capacity~\cite{Kommareddy2020DeACTAV, Ryoo2018ACF, Yan2019NimblePM, Agarwal2017ThermostatAP, Choi2021DancingIT, Lee2022OptimizingTP, hpca24_dynn}. Using the CXL memory, it is natural to build a tiered memory where local DDR is the fast tier and the CXL memory is a slow tier.}
%For instance, the tiered memory systems in data centers can be leveraged to improve the performance and reduce the cost~\cite{} through processes of page allocation, caching, and page migration among different tiered memory. 
\textcolor{revision}{Many solutions~\cite{Maruf2022MULTICLOCKDT, Kumar2021RadiantEP, Hildebrand2020AutoTMAT, Li2022TransparentAL, Ren2021SentinelET, Wang2019PantheraHM, Dulloor2016DataTI, Raybuck2021HeMemST, Kannan2017HeteroOSO, Kim2021ExploringTD, Weiner2022TMOTM, sc18:wu, unimem:sc17, ics21:warpx, wahlgren2023quantitative, ppopp23:merchandiser,eurosys24:mtm,atc24_hm,micro18:pim,8048928,neurips20:hm-ann,ics21:athena,ppopp21:sparta,luo:NGS,hpca25:dlrm,hpca25:buffalo,betty:asplos23,hpdc16:wu} have been proposed to explore and leverage the tiered memory systems to improve application performance. }
%%%%For example, AutoNUMA~\cite{autonuma} is a feature in modern OS, designed to optimize the performance of systems with NUMA architecture.  Tiering-0.8~\cite{tiering0.8} is a recent patch in Linux for memory tiering to improve tiered-AutoNUMA. TPP~\cite{10.1145/3582016.3582063} is a page placement policy for CXL-enabled tiered-memory based on proactive page reclaim. MEMTIS~\cite{lee2023memtis} is a most recent tiered memory system that adopts informed decision-making for page placement and page size determination. 
\textcolor{revision}{However, the performance of those solutions on CXL-ready systems (especially their interplay with page interleaving) is not clear~\cite{Sim2023ComputationalCS, Ryu2023SystemOO, Kim2023SMTSM}.}
\section{Conclusions}
\label{sec:conclusion}
%We characterize the performance of the real CXL memory and explore its use cases. %The emerging CXL memory, although enabling scalable memory bandwidth and capacity, causes longer memory access latency. 
%We explore the use of CXL for LLM training and inference, study the performance of the CXL memory with a spectrum of HPC applications, create a new interleaving policy, and identify the limitations of page migration solutions in the real CXL.

In this paper, we characterize the performance of the real CXL memory and explore its use cases. 
\textcolor{revision}{We explore the use of CXL for LLM training and inference, study the performance of the CXL memory with a spectrum of HPC applications, and investigate how application-transparent solutions (page interleaving and memory tiering) perform with the real CXL. We also create a data object-level interleaving method that switches between the interleaving and NUMA node-preferred polices at the data object level, and demonstrate its superior performance. We hope that our study can shed some lights on how the future HPC systems can leverage CXL memory expansion.}

\begin{comment}
investigate how application-transparent solutions (page interleaving and memory tiering) perform with the real CXL. %\sout{We also create a data object-level interleaving method that switches between the interleaving and NUMA node-preferred polices at the data object level, and demonstrate its superior performance.(unclear sentence).} 
\textcolor{sherry}{We also create a data object-level interleaving method that uses interleaving for specific data objects and NUMA node-preferred policy for others, demonstrating superior performance.}
\end{comment}

%Our study reveals how to interleave CXL and other NUMA memory nodes for high performance. 

%In this paper, we have taken a first step to analyze the device- specific characteristics of true CXL memory and compared them with NUMA-based emulations, a common practice in CXL research. Our analysis revealed key differences between emulated and true CXL memory, with important performance implications. Our analy- sis also identified opportunities to effectively use CXL memory as a memory bandwidth expander for memory-bandwidth-intensive applications, which leads to the development of a CXL-memory-aware dynamic page allocation policy and demonstrated its efficacy.
\section*{Acknowledgment}

This work was partially supported by U.S. National Science Foundation (2104116, 2316202 and 2348350) and the Chameleon Cloud. We thank the anonymous reviewers, as well as our shepherd, for their feedback on the paper.

%%%%%%% -- PAPER CONTENT ENDS -- %%%%%%%%

%%%%%%%%% -- BIB STYLE AND FILE -- %%%%%%%%
\bibliographystyle{IEEEtranS}
\bibliography{li,yang,jianbo,jieliu,ren, xi_sherry}
%%%%%%%%%%%%%%%%%%%%%%%%%%%%%%%%%%%%

\end{document}